\documentclass[showpacs,prl,onecolumn,aps,superscriptaddress,preprintnumbers,nofootinbib]{revtex4}
\usepackage[T1]{fontenc}
\usepackage[latin9]{inputenc}
\usepackage{amsmath,amssymb}
\usepackage{epsfig}
\usepackage{graphicx}
\usepackage{amsmath}
\usepackage{amsfonts}
\def\slashchar#1{\setbox0=\hbox{$#1$}     		
   \dimen0=\wd0                                 	
   \setbox1=\hbox{/} \dimen1=\wd1               	
   \ifdim\dimen0>\dimen1                        	
      \rlap{\hbox to \dimen0{\hfil/\hfil}}      	
      #1                                        	
   \else                                        	
      \rlap{\hbox to \dimen1{\hfil$#1$\hfil}}   	
      /                                         	
   \fi}

\renewcommand{\vec}{\boldsymbol}
\newcommand{\beq}{\begin{equation}}
\newcommand{\eeq}{\end{equation}}
\newcommand{\bea}{\begin{eqnarray}}
\newcommand{\eea}{\end{eqnarray}}
\newcommand{\baa}{\begin{array}}
\newcommand{\eaa}{\end{array}}

\def\eq#1{{Eq.~(\ref{#1})}}
\def\fig#1{{Fig.~\ref{#1}}}

\newcommand{\bas}{\bar{\alpha}_S}
\newcommand{\as}{\alpha_S}

\newcommand{\nn}{\nonumber}

\newcommand{\rv}{\vec{r}}
\newcommand{\bv}{\vec{b}}

\newcommand{\h}{\frac{1}{2}}
\newcommand{\qu}{\frac{1}{4}}

\newcommand{\kv}{\vec{k}}
\newcommand{\Qv}{\vec{Q}_T}

\newcommand{\Lb}{\left(}
\newcommand{\Rb}{\right)}
\def\pom{{I\!\!P}}

\renewcommand{\vec}[1]{\boldsymbol{#1}}

\def\pom{{I\!\!P}}

\begin{document}
\title{ Energy evolution and the Bose-Einstein enhancement for double
gluon densities}
\author{ E.~ Gotsman}
\email{gotsman@post.tau.ac.il}
\affiliation{Department of Particle Physics, School of Physics and Astronomy,
Raymond and Beverly Sackler
 Faculty of Exact Science, Tel Aviv University, Tel Aviv, 69978, Israel}
 \author{ E.~ Levin}
 \email{leving@post.tau.ac.il, eugeny.levin@usm.cl} \affiliation{Department of Particle Physics, School of Physics and Astronomy,
Raymond and Beverly Sackler
 Faculty of Exact Science, Tel Aviv University, Tel Aviv, 69978, Israel} 
 \affiliation{Departemento de F\'isica, Universidad T\'ecnica Federico Santa Mar\'ia, and Centro Cient\'ifico-\\
Tecnol\'ogico de Valpara\'iso, Avda. Espana 1680, Casilla 110-V, Valpara\'iso, Chile} 

\date{\today}


\begin{abstract}
In this paper we show  that  Bose-Einstein enhancement generates
  strong correlations  which  in the BFKL evolution, increase with energy.
 This increase   leads to   double gluon densities
 ( $\Phi$), which are much larger than the product of the single
 gluon densities ($\phi$). However, numerically,  it turns out 
that the ratio $\Phi/\phi^2 \propto \Lb 1/x\Rb^{\delta_2}$ with
 $\delta_2 \sim \bas/\Lb N^2_c - 1\Rb^{2/3}\,\,\ll\,\,1$, and so we
 do not expect a large correction for the experimental accessible 
range of energies.
 However, for $N_c=3$, $\delta_2 = 0.07 \Delta_{\rm 
BFKL}$ ,
 where $\Delta_{\rm BFKL}$ denotes the intercept of the BFKL Pomeron, and 
 thus we can
 anticipate a substantial increase for  the range of rapidities $Y \sim 20$.
 We show that all $1/(N^2_c -1)$ corrections to the double gluon
 densities stem from  Bose-Einstein enhancement.
\end{abstract}

\preprint{TAUP - 3032/18}

\maketitle

\tableofcontents

\flushbottom

\section{Introduction}


For a long time the double parton distribution functions (DPDFs) and 
 their DGLAP evolution\footnote{ Dokshitzer, Gribov, Lipatov, Altarelli
 and Parisi  (DGLAP)  equation\cite{DGLAP} describes the evolution in
 $\ln(Q^2)$, where  $Q$ is the  hardest transverse momentum in the 
process,
 assuming that $\bas \ln(Q^2) \sim 1 $ but $ \bas \ll 1 $ and $\bas \,\ln(1/x)
 \ll 1$. This evolution was generalized for double parton distributions in 
Refs.\cite{Shelest:1982dg,Zinovev:1982be,Snigirev:2003cq,Korotkikh:2004bz,
Gaunt:2009re,Blok:2010ge}} have been of interest  to  the theoretical 
high energy
 community, and has been discussed in  detail
 \cite{Shelest:1982dg,Zinovev:1982be,Snigirev:2003cq,Korotkikh:2004bz,Gaunt:2009re,Blok:2010ge,Ellis:1982cd,Bukhvostov:1985rn,LLS,LALE,
Ceccopieri:2010kg,Diehl:2011tt,
Gaunt:2011xd,Ryskin:2011kk,Bartels:2011qi,Blok:2011bu,Diehl:2011yj,Luszczak:2011zp,
Manohar:2012jr,Ryskin:2012qx,Gaunt:2012dd,Blok:2013bpa,
vanHameren:2014ava, Maciula:2014pla, Snigirev:2014eua,  Golec-Biernat:2014nsa, Gaunt:2014rua, 
Harland-Lang:2014efa, Blok:2014rza, Maciula:2015vza,Diehl:2015bca}. 
On the
 other hand the BFKL evolution\footnote{ The Balitsky, Fadin, Kuraev and
 Lipatov (BFKL) equation\cite{BFKL} is written for evolution in $x$
 (the energy scale of the process) assuming that $\bas \,\ln(1/x) \sim 1$
 but $\bas \,\ll\,1$ and $\bas \ln (Q^2)\,\ll\,1$. 
This evolution has been considered in  Ref.\cite{MUPA,PESCH,LELU1,
LELU2,LELULOOP} for the double gluon densities.}
 and related to them the double gluon densities  (transverse momentum
 distributions(TMD2s))  have attracted less interest of the theorists,
 in spite of the fact, that they give  the simplest way to estimate the
 possible correlations in the  QCD parton cascade at high energies, where
  experimental observations of the double parton
 interactions 
\cite{Akesson:1986iv,Abe:1997bp,Abe:1997xk,Abazov:2009gc,Aad:2013bjm,
Chatrchyan:2013xxa,Aad:2014rua}  
were made.
 
  In this paper we re-visit the evolution equation in the BFKL
 kinematic region  of  small $x$, where  partons are either gluons or 
 colorless dipoles. In the coordinate representation we use  colorless
 dipoles as  partons, while in the momentum representation, it is more
 convenient to discuss the parton cascade in terms of gluons.  This
 evolution equation was written in
 Ref.\cite{MUPA}(see also Refs.\cite{PESCH}) for the double
gluon densities  $\Phi\Lb x_1, p_{1,T};  x_2,  p_{2,T}\Rb \equiv \Phi\Lb Y
 - y_1, p_{1,T}; Y - y_2,  p_{2,T}\Rb$  with respect to rapidity ($ Y $)
 of the initial hadron (projectile).   Note  that we use notation 
 $\Phi\Lb x_1, p_{1,T};  x_2,  
p_{2,T}\Rb$ for the double gluon density,
and $\phi\Lb x, p_T\Rb$ for the single gluon density. $x_i$
 is the fraction of energy of the gluon `i' while $p_{i,T}$
 denotes its transverse momentum.  In the reference frame where the
 initial hadron is fast moving  $\ln(1/x_i) = Y - y_i$, 
 $Y$ denotes the rapidity of the projectile(hadron) and $y_i$ 
 the rapidity of the parton $i$.  This evolution answers
 the question, what are  the multiplicities  of two colorless
 dipoles in one dipole, that moves with rapidity $Y$. We believe
 that in the spirit of the BFKL evolution we need to answer a different
 question, what is the multiplicity of two gluons with rapidities
 $y_1$ and $y_2$, if we know their multiplicities at $y_1 = y^0_1$
 and $y_2 = y^0_2$. Therefore, the first goal of our paper is re-write
 the evolution equation in a convenient form to answer this question.
 It turns out that such evolution has been  discussed in
 Refs.\cite{LELU1,LELU2,LELULOOP} in the framework of the
 CGC approach (see Ref.\cite{KOLEB} for the review of this
 approach). The evolution equations have been derived in 
these papers in the  desired form, for $y_1 = y_2$. It was shown,
 that  it is not necessary to take into account the non-linear 
corrections
 for the double (and multi) gluons densities, and that the evolution 
equations
  reduce to the BFKL evolution. In Ref.\cite{LELULOOP} this evolution
 is written taking into account the corrections of the order of $1/N^2_c$,
 where $N_c$ is the number of colors.
  The linear  evolution
 equations for the double gluon density are closely related to the BKP 
equations\cite{BKP} for the scattering amplitude, and $1/N^2_c$ 
corrections to
 these equations have been discussed in Refs.\cite{BART1,BART2,LET2,BAWU,BAEW,
AAKLL,AKLL}.

The second goal is to include   the Bose-Einsten 
enhancement
 coming from the correlations of  identical gluons, in the evolution. 
Bose-Einstein
 correlations have drawn  considerable attention recently, since they
 give essential contributions to the azimuthal angle
 correlations\cite{BEC1,BEC2,BEC3,BEC4,BEC5,BEC6}.  
It has been shown (see Ref.\cite{BEC4} for example) that the
 Bose-Einstein enhancement leads to a significant contribution
 to the measured angle correlations. We believe that this fact
 calls for a generalization  of the evolution equation by taking
 into account this enhancement. We show that the Bose-Einstein
 enhancement is responsible for a term in the linear evolution
 which is suppressed as $1/\Lb N^2_c - 1\Rb$, which  has 
been found in Ref.\cite{LELULOOP}. In other words, we state that
 all corrections of the order of $1/N^2_c$ in the evolution equations
 for the double gluon density, stem from the Bose-Einstein enhancement.
  In particular,   the symmetry between the azimuthal
 angle $\varphi$ and  the angle $\pi - \varphi$  does not appear. 
This symmetry, which 
is not based on the principle features of our  CGC approach, reveals
 itself in the scattering amplitudes, but we do not find any indication 
of it in the double gluon densities. We wish to stress that, the
 Bose-Einstein   corrections are closely related to the BKP
 equations\cite{BKP} and generate
 the
 energy behaviour  of the twist four operator which increases
 as $s^{\Delta_4}$,  with    $\Delta_4 > 2 \Delta_2$, 
 $\Delta_2$ denotes the intercept of the BFKL Pomeron (see
  Refs.\cite{BART1,BART2,LET2}). 
 
We have discussed the Bose-Einstein enhancement for the DGLAP
 evolution \cite{GOLELAST},
and have shown that it changes considerably, the high energy
 behavior of the DPDFs.
In particular it turns out that 
  the widely used assumption:
\beq \label{I2}
\Phi\Lb x_1, \vec{p}_{1,T} + \h \vec{q}_T; x_2, \vec{p}_{2,T} - \h \vec{q}_T\Rb\,=\,F\Lb q_T\Rb \,\rho\Lb x_1,x_2\Rb\,\phi\Lb x_1,p_{1,T}\Rb\,\phi\Lb x_2,p_{3,T}\Rb
\eeq
does not hold, even at small $x_1$ and $x_2$, due the Bose-Einstein 
correlations
\footnote{In Ref.\cite{GBS2,GBS} it  is shown that \eq{I2}   is 
a good approximation to 
the
 solution of the  DGLAP  evolution equation for the double parton 
 distribution functions at small $x$ and large $p_T$, but this claim
 is only correct,  when neglecting the contributions of the order of
 $1/(N^2_c - 1)$.}.

Before describing the structure of the paper we would like to
 introduce the observables of interest: the parton
 density $\phi\Lb x, p_t, q_T\Rb$ and the double parton density
 $\Phi\Lb x_1, \vec{p}_{1,T} ; x_2, p_{2,T}; \vec{q}_T\Rb$. The
 single  gluon  density characterizes the multiplicity of 
gluons with  fraction of energy $x$ and transverse momentum
 $\vec{p}_T$  at $q_T = 0$, and it can be written as follows

\bea \label{ISP}
\phi\Lb x, \vec{p}_{T}, \vec{q}_T; b \Rb\,&=& \,\,\sum^\infty_{n=1}\,\int\,\prod^n_{i=1}
\frac{d x_i}{ x_i}\, d^2 k_{i,T}\,\sum_{c_i}\,\langle \Omega^+_n| \Psi^*\Lb \{x_i,\kv_{i,T}; c_i\}\Rb\nn\\
& &\,\Bigg\{a^+(x, \vec{p}_{T} + \h \vec{q}_T; b)\,\,\,a(x, \vec{p}_{T} - \h \vec{q}_T; b)\Bigg\}\,\Psi\Lb \{x_i, \kv_{i,T};c_i\},  \Rb|\Omega_n\rangle
\eea

  where $\Psi$ denotes the partonic wave function of the fast hadron,
   $| \Omega_n \rangle = \prod^n_i \,a^+( x_i, \kv_{i,T},c_i) |0\rangle$
 ($|0\rangle $  denotes  the  vacuum state)
 and  $a^+( x_i, \kv_{i,T},c_i)$ and $a( x_i, \kv_{i,T},c_i)$ 
 denote the creation and annihilation operators 
for  partons (gluons for small $x_i$) with fraction of  energy $x_i$,
 transverse momentum $\kv_{i,T}$ and  color $c_i$.
  The produced gluon has longitudinal momentum $x$ and transverse momentum
 $\vec{p}_{T}$, while  $b$ indicates its color. 
 
The double transverse momentum densities describe  the number of 
gluons with 
 $(x_1,p_{1,T})$ and $(x_2,p_{2,T})$   in the parton cascade, and  it 
can
 be written  with the aid of the wave function of the produced gluon
$\Psi\Lb 
\{x_i,k_{i,T}\}\Rb$ as follows 

\bea \label{I1}
&&\Phi\Lb x_1, \vec{p}_{1,T} ; x_2, \vec{p}_{2,T} ; \vec{q}_T; b, c \Rb\,= \,\,\sum^\infty_{n=2}\,\int\,\prod^n_{i=1}
\frac{d x_i}{ x_i}\,  d^2 k_{i,T}\,\sum_{c_i}\,\langle \Omega^+_n| \Psi^*\Lb \{x_i,\kv_{i,T}; c_i\}\Rb\nn\\
&&\,\Bigg\{a^+(x_1, \vec{p}_{1,T} + \h \vec{q}_T; b)\,a^+(x_2, \vec{p}_{2,T} - \h \vec{q}_T; c)\,a(x_2, \vec{p}_{2,T} + \h \vec{q}_T; c)\,a(x_1, \vec{p}_{1,T} - \h \vec{q}_T; b)\Bigg\}\Psi\Lb \{x_i,\kv_{i,T}; c_i\}\Rb |\Omega_n\rangle\eea

The paper is organized as follows: In the next section we discuss
 the BFKL evolution of the double gluon density in the region of
 low $x$.   For  completeness of presentation we review the
 derivation of the evolution equations for the double gluon 
 densities which was done in Refs.\cite{LELU1,LELU2,LELULOOP}
 in the framework of the dipole approach\cite{MUDI}. In spite
of the fact  that
 this derivation is only valid for $y_1=y_2$, it shows that these 
evolution equations are linear BFKL equations which are not affected 
by   non-linear shadowing corrections. In section 3 we re-derive
 the BFKL equations for the double gluon densities directly in the
 momentum representation. In this representation,  we generalize the
 equation for the case of $y_1 \neq y_2$ and re-write the equations
 in the form which is suitable  for taking into account the Bose-Einstein
 enhancement (BEE). In this section we find the solutions to the
 equations without BEE. The interference diagrams that are
 responsible for the BEE, are discussed in section 4. In
 section 5 the evolution equations with BEE are proposed,
 and   we find that  $\Phi/\phi^2 \propto \Lb 1/x\Rb^{\delta_2}$
 with $\delta_2 \sim \bas/\Lb N^2_c - 1\Rb^{2/3}$. Section 5 is
 devoted to a discussion of the energy behavior of the double gluon
 densities. In the conclusions  we discuss our main results.


\section{BFKL evolution of double dipole densities in the CGC approach.}
In this section we discuss the evolution equation for the double gluon 
densities in the framework of the CGC approach. As  mentioned this
 equation has been derived in Refs.\cite{LELU1,LELU2,LELULOOP}, here we 
give a brief review both for completeness of the presentation, and to 
 display the main features of the multi gluon densities, which are 
not the 
main subject of these papers. The partonic wave function can be expanded 
as the sum of Fock states with fixed multiplicity of partons (colorless 
dipoles):
\beq \label{CGC1}
\Psi\Big( \{\vec{r}_i,\vec{b}_i\}\Big)\,\,=\,\,\sum_{n}\,\alpha_n |\Omega_n\rangle
\eeq
The colorless dipole is characterized by two variables: its size
 $\vec{r}_i$ and its impact parameter $\vec{b}_i$. However, in
 this paper we will  sometimes use a different set of  
variables: $\vec{x}_i$ for the position of the quark and $\vec{y}_i$
 for the position of the anti-quark in the dipole. One can see
 that $\vec{r}_i \,=\vec{x}_i \,-\,\vec{y}_i$ and
 $\vec{b}_i \,=\,\h\Lb \vec{x}_i \,+\,\vec{y}_i\Rb$. $\alpha_n^2$
 is the probability to find $n$ dipoles with the same value of rapidity $Y$ : 
\beq \label{CGC2}
\alpha^2_n\,=\,P_n\Lb Y; \{\vec{r}_i,\vec{b}_i\}\Rb
\eeq
 The QCD cascade can be written as the linear
 functional equation for the following functional\cite{MUDI,LELU1}:
\beq \label{CGC3}
Z\left(\tilde{Y} = Y - y;\,[u_i] \right)\,\,\equiv\,\,\sum_{n=1}\,\int\,\,
P_n\left(Y\,;\,\{\vec{r}_i,\vec{b}_i\}
 \right) \,\,
\prod^{n}_{i=1}\,u(r_i, b_i) \,d^2\,r_i\,d^2\,b_i
\eeq 
where $u(r_i, b_i) \equiv u_i $ is an arbitrary function of $r
_i$ and $b_i$ and $y = y_1 = y_2$.
It  follows immediately from \eq{CGC2}
that the functional obeys the condition:
at $u_i\,=\,1$ 
\beq \label{CGC4} 
Z\left(\tilde{Y}\,;\,[u_i=1]\right)\,\,=\,\,1\,.
\eeq
The physical meaning of  \eq{CGC4} is that the sum over
all probabilities is one.

~

\subsection{Balitsky-Kovchegov parton cascade}

 To write the evolution equation,
we need to specify the QCD processes with  color dipoles. 
  The parton cascade that leads to the Balitsky-Kovchegov
 non-linear equation for the scattering amplitude, stems from the process 
of 
the decay of one dipole to two dipoles, which gives the main contribution 
at
 the leading order of perturbative QCD at large $N_c$\cite{MUDI}. The
 probability of this decay is equal to
\beq \label{BKC1}
P_{1 \to 2}\Lb |\vec{r}_1 + \vec{r}_2| \to r_1 + r_2 \Rb\,\,=\,\,\frac{\bas}{2 \,\pi}\,\frac{\Lb\vec{r}_1 +\vec{r}_2\Rb^2 }{r^2_1\,r^2_2}
\eeq

Bearing \eq{BKC1} in mind, we can write the linear equation for $Z$:
\beq \label{LEQZ}
\,\frac{\partial \,Z}{\bas\,\partial \,\tilde{Y}}\,\,=\,\,-\,\,
\int\,d^2 r\,d^2 b\,\,V_{1\rightarrow 1}(\rv ,\,\bv,\,[u])\,\, Z\,\,
+\,\,\int\,\,d^2 \,r \,d^2\,r' \,d^2 b\,\,V_{1\rightarrow 2}(\rv ,\,\rv', \,\bv,\,[u])\,\, Z\,
\eeq
with the definitions
\beq \label{V11} 
V_{1 \rightarrow 1}(\rv ,\, \bv,\,[u])\,\,=\,\,
\,\,\omega(r) \,\,u(\vec{r},\,\vec{b})\,\,\frac{\delta}{\delta u(\vec{r},\vec{b})}~\mbox{with}~
\omega\Lb r\Rb\,\,=\,\,\int d^2 r' \,P_{1 \to 2}\Lb r \to r' +  |\vec{r} - \vec{r}'| \Rb\,\,=\,\,\frac{\bas}{2 \pi}\int d^2 r'\frac{r^2}{r'^2\,\Lb \vec{r} - \vec{r}'\Rb^2}
\eeq
and
\beq \label{V12}
V_{1 \rightarrow 2}(\rv,\,\rv',\,\bv,\,[u])\,\,
=\,\,\frac{ \bas}{2\,\pi} 
\,\,\frac{r^2}{r'^2\,(\vec{r} -\vec{ r}')^2}\,\,
\,u\Lb\vec{r}', \,\vec{b}\,+\,\h(\rv\,-\,\rv')\Rb\,\,u\Lb\rv \,-\,\rv', \,\bv\,-\,\h \rv\Rb\,\,
\frac{\delta}{\delta u(\rv, \,\bv)}\,.
\eeq 
The functional derivative with respect to $u(r,b)$,  plays the  role 
of an  annihilation operator for a dipole of  size $r$,  at  impact 
parameter $b$. 
The multiplication by $u(r,b)$ corresponds to
a creation operator for this dipole. Therefore,\eq{LEQZ} is a typical
 cascade  equation in which the first term describes the depletion of
 the probability due to splitting into $n+1$ dipoles, while the second
 term is responsible for the growth due to splitting of $(n-1)$ dipoles
 into $n$ dipoles. From \eq{CGC3}, one can see that the multi dipole 
density 
$\rho^{(n)}\Lb Y - Y_0; \{ \rv_i,\bv_i\}\Rb$ can be found as follows
\bea \label{RON}
\rho^{(n)}\Lb Y - y; \{ \rv_i,\bv_i\}\Rb\,&=&\,\frac{1}{n!}\,\prod^n_{i =1}
\,\frac{\delta}{\delta
u_i } \,Z\left(Y\,-\,y;\,[u_i] \right)|_{u=1}\,\,\\
&=&\,\,\sum^\infty_{k=n}\frac{k!}{(k - n)!\,n!}\,\int P_k\left(Y - y\,;\,\vec{r}_1,\vec{b}_1, \dots,\vec{r}_n,\vec{b}_n,\{ \{\vec{r}_i,\vec{b}_i\}
 \right) \,\,
\prod^{k}_{i= n+1}\, \,d^2\,r_i\,d^2\,b_i\nn
\eea
which gives the  probability of finding  $n$-dipoles with the given 
kinematics.

From   \eq{LEQZ} we obtain
\bea\label{EVRON}
&&\frac{\partial \,\rho^{(n)}(\tilde{Y} ; \rv_1, \bv_1\,\ldots\,,\rv_n, \bv_n)}{ 
\bas\,\partial\,\tilde{Y}}\,\,=\,\,
-\,\sum_{i=1}^n
 \,\,\omega(r_i)\,\,\rho^{(n)}( Y - y; \rv_1, \bv_1\,\ldots\,,\rv_n, \bv_n)\,\,\\
 &&+
2\,\sum_{i=1}^n\,
\int\,\frac{d^2\,r'}{2\,\pi}\,
\frac{r'^2}{r^2_i\,(\rv_i\,-\,\rv')^2}\,
\rho^{(n)}(Y - y; \ldots\,\rv', \bv_i- \h\rv', \dots) 
+\sum_{i=1}^{n-1}\,\frac{(\rv_i + \rv_n)^2}
{(2\,\pi)\,r^2_i\,r^2_n}\,
\rho^{(n-1)}(Y - y; \dots,\rv_i\,+\,\rv_n,\bv_{in},\dots)\nn
 \eea
For $\rho^{(2)}$ we have
\bea\label{EVRO2}
&&\frac{\partial \,\rho^{(2)}(\tilde{Y}; \rv_1, \bv_1;\rv_2, \bv_2)}{ 
\bas\,\partial\,\tilde{Y}}\,\,=\,\,
-\,\sum_{i=1}^2
 \,\,\omega(r_i)\,\,\rho^{(2)}( \tilde{Y} ; \rv_1, \bv_1;\rv_2, \bv_2)\,\,\\
 &&+
2\,\Bigg(\,
\int\,\frac{d^2\,r'}{2\,\pi}\,
\frac{r'^2}{r^2_1\,(\rv_1\,-\,\rv')^2}\,
\rho^{(2)}(\tilde{Y}; \rv', \bv_1- \h\rv',\rv_{2},\bv_{2} \dots)\, +\,\Big( 1\, \leftrightarrows\, 2\Big) \Bigg) \,\,+\,\,\frac{(\rv_1 + \rv_2)^2}
{(2\,\pi)\,r^2_1\,r^2_2}\,
\rho^{(1)}(\tilde{Y}; \rv_i\,+\,\rv_2, \bv_{12})\nn
 \eea
There are two main features of the equation: (i) there are no non-linear
 corrections, and (ii) we have two contributions: BFKL evolution of 
$\rho^{(2)}$
 and  the contribution to $\rho^{(2)}$ from  single parton showers.
 This structure is the same as in the DGLAP evolution
 (see Refs.\cite{Shelest:1982dg,Zinovev:1982be,Snigirev:2003cq,Korotkikh
:2004bz,Gaunt:2009re,Blok:2010ge}).

\begin{boldmath}
\subsection{$1/N^2_c$ corrections to the Balitsky-Kovchegov cascade}
\end{boldmath}

  \begin{figure}[h]
    \centering
  \leavevmode
      \includegraphics[width=10cm,]{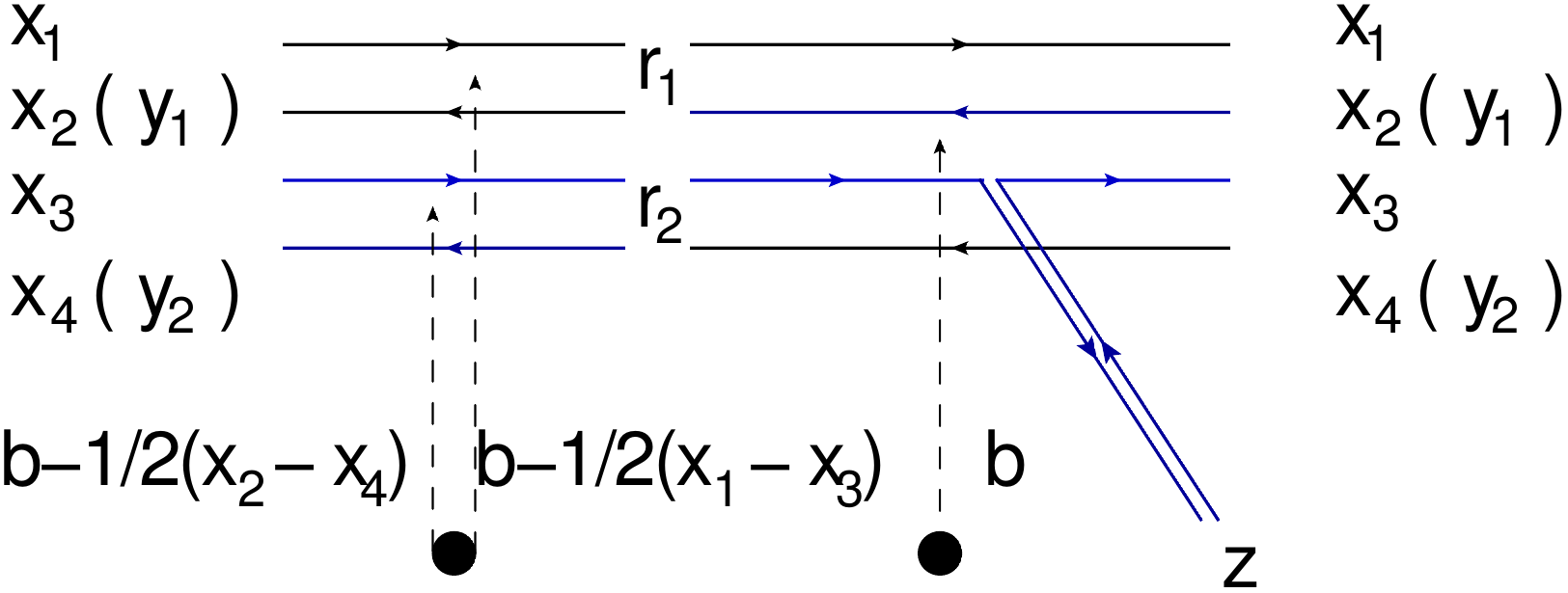}  
    \caption{A graphical representation of the decay of two dipoles to 
three dipoles:
 $x_{12} + x_{34}\,\to\,x_{23} + x_{14} \,\to x_{z2} \,+\,x_{3z} + x_{14}$.
 The lines of the same colors indicate  the colorless dipole which decays 
in
 two dipoles due to emission of a gluon with  coordinate $\vec{z}$.
 }
\label{23}
  \end{figure}

In Ref.\cite{LELULOOP} it is suggested to add  the following term to \eq{LEQZ}:
\bea \label{V23}
&&\int \prod_{i=1}^2 d^2 x_i \,d^2 y_i\,d^2 z\,\,\frac{1}{N^2_c - 1} P_{1 \to
 2}\Lb| \vec{x}_2 - \vec{y}_1| \to \,| \vec{x}_2 - \vec{z}|\,+\,| \vec{z} -
 \vec{y}_1|\Rb\nn\\
&&\times\,\,\Big( 1 - u\Lb \vec{x}_1,\vec{y}_2\Rb\Big)\,\Big( u\Lb \vec{x}_2,
\vec{y}_1\Rb\,-\,u\Lb \vec{z},\vec{y}_1\Rb\, u\Lb \vec{x}_2,\vec{z}\Rb\Big)
 \frac{\delta}{\delta u\Lb \vec{x}_1,\vec{y}_1\Rb}\,\frac{\delta}{\delta u\Lb
 \vec{x}_2,\vec{y}_2\Rb}\,Z\Lb Y, [ u_i]\Rb
\eea
The process that is described by \eq{V23} is shown in \fig{23}. Indeed, if
 two dipoles consist of   quarks and antiquarks with the same colors, 
 the dipole, in which one quark from one dipole and the antiquark from
 another dipole, can create a different dipole which decays into $q, 
\bar{q}$ and a gluon.
The term of \eq{V23} generates the$1/(N^2_c-1)$  corrections to \eq{EVRO2}
 which have the following form:
\bea \label{EVRO2NC}
&&\hspace{-0.5cm}\frac{1}{N^2_c - 1}\Bigg\{ P_{1 \to 2}\Big( (\vec{x}_1, \vec{y}_2) \to (\vec{x}_1, \vec{y}_1)\,+\,(\vec{x}_2, \vec{y}_1)\Big) \,\delta^{(2)}\Lb \vec{x}_2 - \vec{y}_1\Rb\,\rho^{(1)}\Lb\tilde Y;  \vec{x}_2, \vec{y}_2\Rb\,+\, P_{1 \to 2}\Big( (\vec{x}_2, \vec{y}_1) \to (\vec{z}, \vec{y}_1) +(\vec{x}_2, \vec{z})\Big)\nn\\
&&\times\,\,\Bigg(\rho^{(2)}\Lb\tilde{Y};  \vec{x}_1, \vec{y}_2;
\vec{z}, \vec{y}_1\Rb\,+\,\rho^{(2)}\Lb \tilde{Y};\vec{x}_1, \vec{y}_2;
\vec{x}_2, \vec{z }\Rb\,\,-\,\, \rho^{(2)}\Lb\tilde{Y} ;\vec{x}_1, \vec{y}_2;
\vec{x}_2, \vec{y}_1\Rb \Bigg) +\Big( 1\, \leftrightarrows\, 2\Big)\Bigg\}
\eea

One can see that the structure of \eq{EVRO2NC}  is similar to that of
\eq{EVRO2}: i.e.
 the production of two dipoles from the single parton cascade, and 
the evolution of the double density with a kernel which is different
 from \eq{EVRO2}, although it consists of the same elements. We need to 
re-write
 \eq{EVRO2} with additional term of \eq{EVRO2NC}  in the momentum
 representation, to obtain the more familiar form of the double gluon
 density evolution equation. However, we chose a different strategy:
 we derive the equation directly in the momentum representation. First,
 we believe that we can relax the assumption that $y_1=y_2$
and second, we hope that this derivation will  clarify the physical 
meaning
 of the  additional term of \eq{EVRO2NC}.

 We would like to stress that the derivation, which we have
 discussed, is very instructive for understanding  
the different contributions to the evolution due to the clear
 physics  interpretation  of the dipole approach in perturbative QCD.

~

\section{BFKL evolution  without  Bose-Einstein enhancement}
\begin{boldmath}
\subsection{Equations for $\bas |y_2  - y_1| \,\gg\,\,1$.}\end{boldmath}

In this section we re-write \eq{EVRO2} in the momentum representation using
  two lessons from the derivation of the previous section. The double parton
 density is not affected by the shadowing (screening) corrections and obeys
 the linear BFKL  equations, and the equation should match \eq{EVRO2} at 
 $y_1=y_2$.

  In the BFKL  region we consider that $\bas  \,\ln\Lb1/x_i\Rb  \,=\,Y - y_i
 \,\gg\,1$  while $ \bas \ln(p_{i,T}^2/Q^2_0) \,\ll\,1$, and  $\bas \,\ll
 \,1$ and the evolution equations sum the contributions of the order of
 $\Lb \bas \ln(1/x)\Rb^n$ (leading log(1/x) approximation (LLA)).
        
  \begin{figure}[ht]
    \centering
  \leavevmode
      \includegraphics[width=14cm,height= 2.5cm]{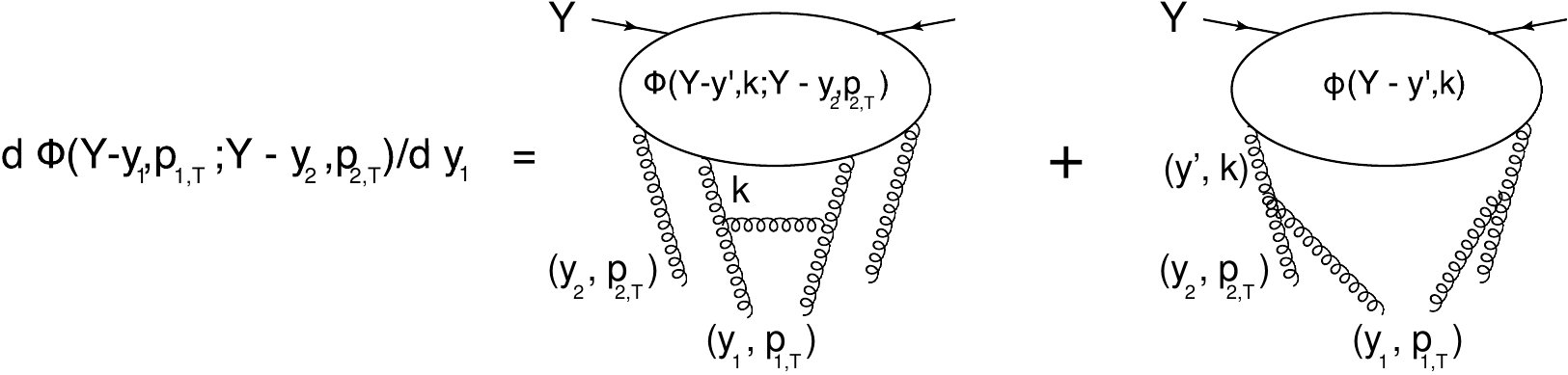}  
    \caption{The graphical representation of the evolution equation(see
 \eq{EQBFKL1} and \eq{EQBFKL2}).}
\label{bfkl}
  \end{figure}

   
 In the region of small $x_i$ only a gluon can be produced \cite{KOLEB},
 and for the double gluon density we  expect to have two equations 
 of the following forms(see \fig{bfkl})
 
\bea 
\frac{\partial \Phi\Lb Y - y_1,p_{1,T}; Y - y_2,p_{2,T}; q_T\Rb}{\partial \,\Lb Y - y_1\Rb} \,\,&=&\,\, \bas \int \frac{d^2 k_T}{2 \pi} K\Lb p_{1,T}, k_T; q_T\Rb \,\Phi\Lb Y - y_1,k_{T}; Y - y_2,p_{2,T}; q_T\Rb\,\nn\\
&+&\,\,\bas\,\phi\Lb Y - y', \vec{p}_{1,T} + \vec{p}_{2,T}\Rb \,\Gamma\Lb p_1,p_2; y_1, y_2\Rb;\label{EQBFKL1}\\
\frac{\partial \Phi\Lb Y - y_1,p_{1,T}; Y - y_2,p_{2,T}; q_T \Rb}{\partial \,\Lb Y - y_2\Rb} \,\,&=&\,\, \bas \int \frac{d^2 k_T}{2 \pi} K\Lb p_{2,T}, k_T; q_T\Rb \,\Phi\Lb Y - y_1,p_{1,T}; Y - y_2, k_T\Rb\,\nn\\
&+&\,\,\bas\,\phi\Lb Y - y', \vec{p}_{1,T} + \vec{p}_{2,T}\Rb \,\Gamma\Lb p_1,p_2; y_2, y_1\Rb;\label{EQBFKL2}
 \eea
 where $K\Lb p_T,k_T\Rb $ is the BFKL kernel which is equal to \cite{BFKL}
 \bea 
 K\Lb p_T,k_T; q_T\Rb\,\,&=&\,\,\frac{1}{\Lb \vec{p}_T\, -\, \vec{k}_T \Rb^2}\Bigg\{ \frac{ \Lb \vec{p}_T - \h \vec{q}_T\Rb^2}{\Lb \vec{k}_T - \h \vec{q}_T\Rb^2}\,+\, \frac{ \Lb \vec{p}_T + \h \vec{q}_T\Rb^2}{\Lb \vec{k}_T + \h \vec{q}_T\Rb^2} \Bigg\} \,-\,\frac{q^2_T}{\Lb \vec{k}_T + \h \vec{q}_T\Rb^2\,\Lb \vec{k}_T - \h \vec{q}_T\Rb^2}\nn\\
  \,\,&-&\,\,\Bigg\{ \omega_G\Lb \vec{p}_T + \h \vec{q}_T\Rb \,+\,\omega_G\Lb \vec{p}_T - \h \vec{q}_T\Rb\Bigg\}\,  
\delta^{(2)}\Lb \vec{p}_{T} - \vec{k}\Rb \label{KER}\\
\omega_G\Lb p_T\Rb &=&\,\,\h \,p^2_T \int \frac{d^2 k_T}{2 \pi} \frac{1}{\Lb \vec{p}_T\, -\, \vec{k}_T \Rb^2\,k^2_T}\label{KEROM} \\
 K\Lb p_T,k_T; q_T= 0\Rb\,\,&=&\,\,\frac{2}{\Lb \vec{p}_T\, -\, \vec{k}_T \Rb^2} \frac{p^2_T}{k^2_T} - 2 \omega_G\Lb p_T\Rb\,\delta^{(2)}\Lb \vec{p}_T - \vec{k}_T\Rb\label{KER0}
\eea

\begin{figure}[ht]
    \centering
    \begin{tabular}{ccc}
  \leavevmode
      \includegraphics[width=6.5cm,height= 3.5cm]{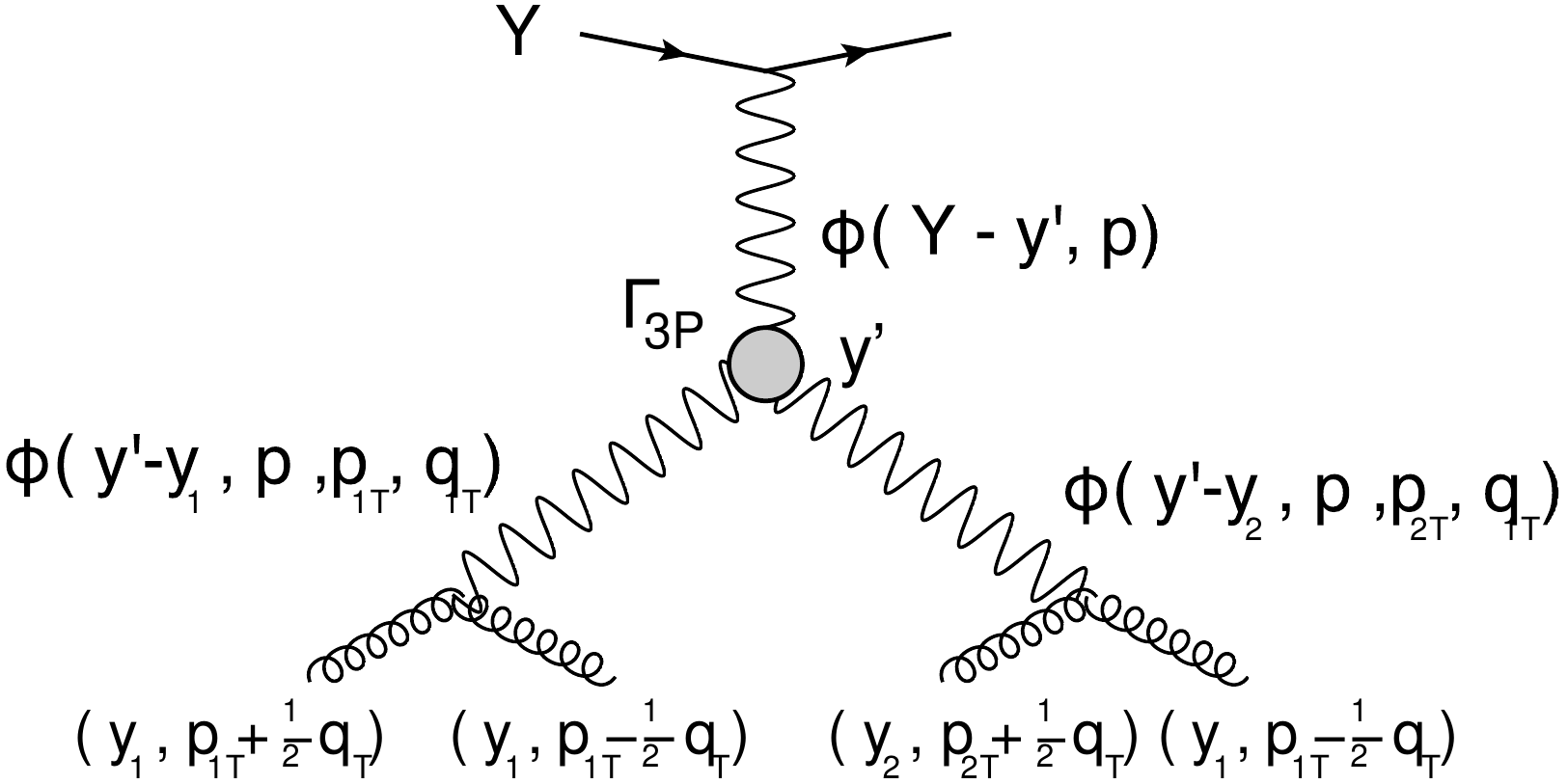}&~~~~~~~  &\includegraphics[width=6.5cm,height= 3.5cm]{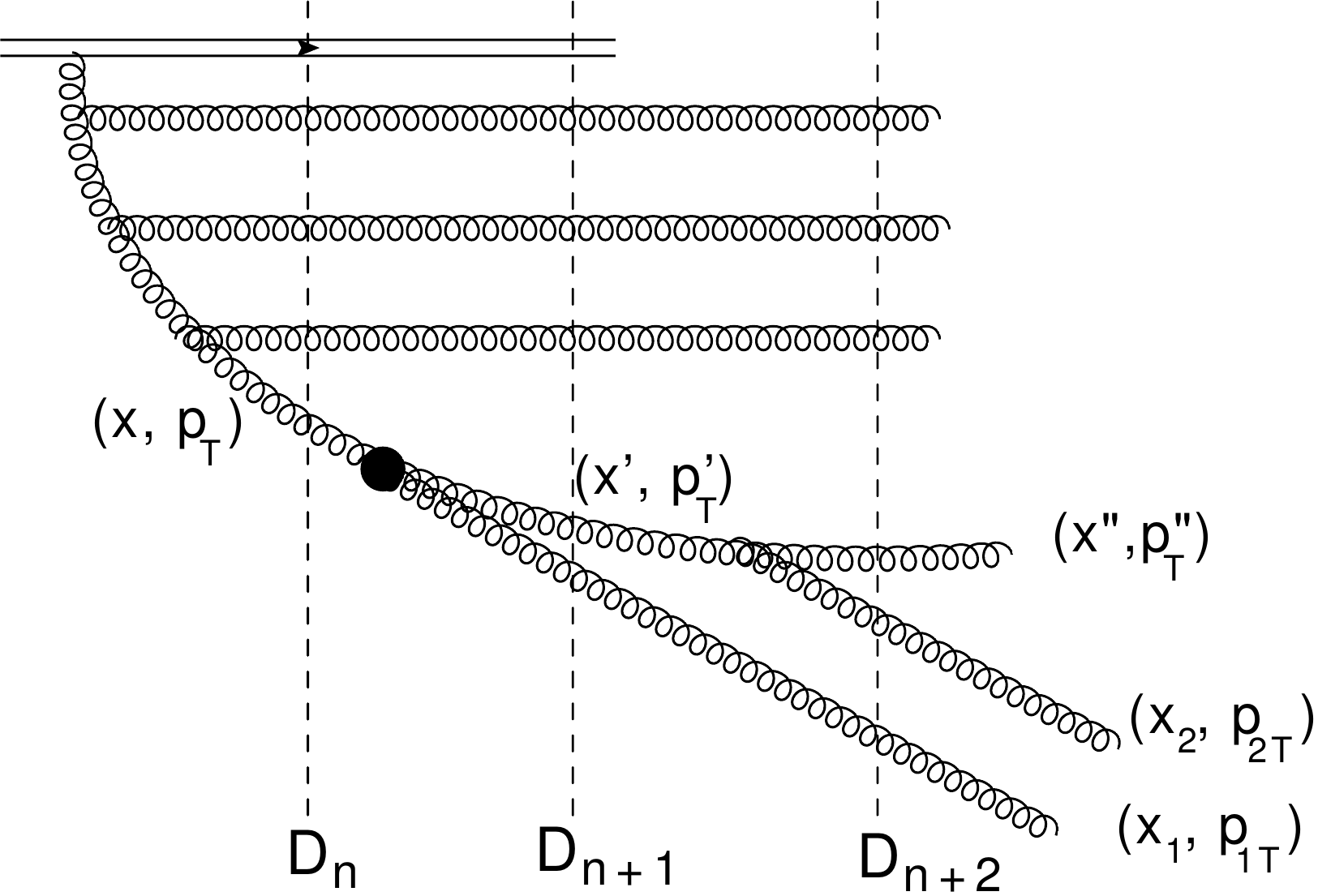} \\
      \fig{bfkl3p}-a & & \fig{bfkl3p}-b\\
      \end{tabular}

        \caption{\fig{bfkl3p}-a:The graphical representation of the
 triple BFKL Pomeron diagrams. For simplicity we show the evolution
 at $q_T = 0$.  \fig{bfkl3p}-b: the structure of the partonic wave 
function in the light-cone perturbative approach  for the production
 of two gluons from the single parton cascade at $x_1 = x_2$ ($y_1 = y_2$).
 We denote  by $D_n$ the dominator of the propagator for the state that has
 $n$-gluons.}
\label{bfkl3p}
  \end{figure}

      The non -homogenous term takes into account the possibility to 
produce two gluons  from a single gluon cascade.
 The expression for these terms  is  written directly from
 the second diagram of  \fig{bfkl}.   Function $\Gamma\Lb \vec{p}_{1T},
\vec{p}_{2T}; y_1,y_2\Rb$ has to be found. It is clear that for
 $\bas |y_2  - y_1| \,\gg\,\,1$ two emitted gluons in \fig{bfkl} with
 rapidities $y_1$ and $y_2$ can emit gluons, and the observed gluons will
 be  amongst them. Therefore, the  general diagram, which determines the
 non-homogenous term is the  triple BFKL Pomeron diagram of \fig{bfkl3p}-a.

 This diagram  can be written in
 the following general form
 {\small
 \beq \label{3P} 
 \Phi_{3\pom}\Lb Y,y_1,y_2; \vec{p}_{1,T}, \vec{p}_{2,T},q_T\Rb\,\,=\,\,\bas \int^Y_{y_2 \,\geq\,y_1}\!\!\!\!\!\!\!\!\!\!\!\!\!d y'\,d^2 p_T\phi_{\rm pr}\Lb Y - y', p_T\Rb \,\Gamma_{3 \pom}\Lb p_T\Rb\,\phi\Lb y' - y_1, \vec{p}_{T},\vec{p}_{1,T},q_T\Rb\,\phi\Lb y' - y_2, \vec{p}_{T},\vec{p}_{2,T},q_T\Rb 
\eeq }
where $\Gamma_{3\pom}$ denotes the triple BFKL Pomeron vertex which we 
will
 discuss below. The single gluon densities in \eq{3P} $\phi_{\rm pr}$ 
and $\phi$ are different, since $\phi_{\rm pr}$ is the density of the
 gluons in the projectile (hadron), while $\phi$ is the density of the
 gluons in the cascade of  the single gluon with rapidity $Y - y'$.  
This term in the evolution  was considered for the first time  in
 Ref.\cite{MUPA}. We refer our readers to this paper or more detail.

 The contributions of this diagram  to the evolution equation have the
 following form for $y_2 > y_1$:
 {\small
 \bea 
1.\,\, \frac{\partial  \Phi_{3\pom}\Lb Y,y_1,y_2; \vec{p}_{1,T}, \vec{p}_{2,T},q_T\Rb}{\partial y_1} &=&\bas \int^Y_{y_2 \,\geq\,y_1}\!\!\!\!\!\!\!\!\!\!\!\!\!\!\!\!d y'\,d^2 p_T\,\phi_{\rm pr}\Lb Y - y', p_T\Rb \,\Gamma_{3 \pom}\Lb p_T\Rb\,\frac{\partial \phi}{\partial y_1}\Lb y' - y_1, \vec{p}_{T},\vec{p}_{1,T},q_T\Rb\,\phi\Lb y' - y_2, \vec{p}_{T},\vec{p}_{2,T},q_T\Rb ; \nn \\
2.\,\,  \frac{\partial  \Phi_{3\pom}\Lb Y,y_1,y_2; \vec{p}_{1,T}, \vec{p}_{2,T},q_T\Rb}{\partial y_2} &=&\bas \int^Y_{y_2 \,\geq\,y_1}\!\!\!\!\!\!\!\!\!\!\!\!\!\!\!\!d y'\,d^2 p_T\,\phi_{\rm pr}\Lb Y - y', p_T\Rb \,\Gamma_{3 \pom}\Lb p_T\Rb\, \phi\Lb y' - y_1, \vec{p}_{T},\vec{p}_{1,T},q_T\Rb\,\frac{\partial \phi}{\partial y_2}\Lb y' - y_2, \vec{p}_{T},\vec{p}_{2,T},q_T\Rb  \nn\\
  &+& \,\bas \,\int d^2 p_T \,\phi_{\rm pr}\Lb Y - y_2, p_T\Rb  \,\Gamma_{3 \pom}\Lb p_T \Rb \phi\Lb y_2  - y_1, \vec{p}_{T},\vec{p}_{1,T},q_T\Rb\,\phi\Lb 0, \vec{p}_{T},\vec{p}_{2,T},q_T\Rb;\label{3PEV}
\eea   }

From \eq{EQBFKL1} and \eq{EQBFKL2}  one can see that the first term in
 the both equations  is included  in the first terms of the evolution 
equations since
\beq \label{3P3}
\frac{\partial \phi}{\partial y_i}\Lb y' - y_i, \vec{p}_{T},\vec{p}_{2,T},
 q_T\Rb\,\,=\,\,- \bas\,\int d^2 k_T\,\,K\Lb p_{i,T},k_T; q_T\Rb\,\,\phi\Lb
  y' - y_i, \vec{p}_{T},\vec{p}_{2,T},q_T\Rb
\eeq

We also see  that \eq{3PEV} does not generate the non-homogenous term in 
\eq{EQBFKL1} if $y_1  \,< \,y_2$,  in the case which we consider
 here: $y_2 > y_1$.  We need to specify the triple BFKL
 Pomeron vertex and  $\phi\Lb 0, \vec{p}_{T},\vec{p}_{2,T},q_T\Rb$ in
 \eq{3PEV}.  Actually 
\beq \label{3P4}
\phi\Lb 0, \vec{p}_{T},\vec{p}_{2,T},q_T\Rb\,\,=\,\,\delta^{(2)}\Lb
 \vec{p}_T  - \vec{p}_{2,T}\Rb
\eeq
 In the diagram of \fig{bfkl3p}-a the triple Pomeron vertex enters 
with the momentum transferred along the upper Pomeron being equal to
 zero. We believe that we can find  this vertex directly from the
 non-linear Balitsky-Kovchegov evolution equation\cite{BK} for the
 scattering dipole amplitude: 
 \bea \label{BK}
  \frac{\partial N\Lb Y; x_{12}, b\Rb}{\partial \,Y}\,\,&=&\,\,\bas \,\int \frac{d^2 x_3}{\pi}\,\frac{x^2_{12}}{x^2_{13}\,x^2_{23}
 } \Big\{N\Lb Y; x_{13}, \vec{b} - \h \vec{x}_{23}\Rb\,+\,
 N\Lb Y; x_{23},
 \vec{b} - \h \vec{x}_{13}\Rb \,-\,N\Lb Y;  x_{12}, b\Rb\,\,\nn\\
 &-&\,\,N\Lb Y; x_{13}, \vec{b} - \h \vec{x}_{23}\Rb\,N\Lb Y; x_{23},
 \vec{b} - \h \vec{x}_{13}\Rb \Big\}
      \eea
 where $x_{ik} \,=\,\vec{x}_i \,-\,\vec{x_k}$ and $b$ denotes the impact 
factor.
 \eq{BK} in the momentum representation, which we define as
 \beq \label{MR}
 N\Lb Y; x_{12}, b\Rb\,\,=\,\,x^2_{12}\int \frac{d^2 k_T}{2 \pi} \,\frac{d^2 q_T}{2 \pi} e^{ i \vec{x}_{12} \cdot \vec{k}_T\,\,+\,\,i\,\vec{b} \cdot \vec{q}_T } \,\phi\Lb Y; k_T,q_T\Rb
 \eeq
 has the following form:
 {\small\beq \label{BKMR}
\frac{\partial  \phi\Lb Y; k_T, q_T = 0 \Rb }{\partial Y}\,\,=\,\,\bas \,\int \frac{d^2 k'_T}{2 \pi} \,K\Lb k_T, k'_T; q_T=0\Rb \,\phi\Lb Y, k'_T, q_T=0\Rb\,\,-\,\,\bas \,\int \frac{ d^2 q_T}{2 \pi}\,\phi\Lb Y; k_T,\vec{q}_T\Rb \,\phi\Lb Y; k_T, -\vec{q}_T\Rb 
\eeq }
   We can build the diagram of \fig{bfkl3p}-a by  iterating
 \eq{BKMR}, using the non-linear term as the first iteration. Returning
 to \eq{3PEV}, we see the the non-homogeneous term has the form:
\bea \label{3PFIN}
&&\bas \int\!\!d^2 p_T \,\phi_{\rm pr}\Lb Y - y_2, p_T\Rb  \,\Gamma_{3 \pom}\Lb p_T \Rb \phi\Lb y_2  - y_1, \vec{p}_{T},\vec{p}_{1,T},q_T\Rb\,\phi\Lb 0, \vec{p}_{T},\vec{p}_{2,T},q_T\Rb\,\,= \\
 &&\bas \, \phi_{\rm pr}\Lb Y - y_2, p_{2,T}\Rb\, \phi\Lb y_2  - y_1, 
\vec{p}_{2,T},\vec{p}_{1,T},q_T\Rb\nn
 \eea 
 Finally, the set of evolution equations can be re-written in the following
 formf or $\bas | y_1 - y_2| \,gg\,1$ :
 \bea 
\frac{\partial \Phi\Lb Y - y_1,p_{1,T}; Y - y_2,p_{2,T}; q_T\Rb}{\partial \,\Lb Y - y_1\Rb} \,\,&=&\,\, \bas \int \frac{d^2 k_T}{2 \pi} K\Lb p_{1,T}, k_T; q_T\Rb \,\Phi\Lb Y - y_1,k_{T}; Y - y_2,p_{2,T}; q_T\Rb\,\nn\\
&+&\,\,\bas \,\, \phi_{\rm pr}\Lb Y - y_2, p_{2,T}\Rb\, \phi\Lb y_1  - y_2, \vec{p}_{2,T},\vec{p}_{1,T},q_T\Rb\vartheta\Lb y_1 - y_2\Rb;\label{EQBFKL21}\\
\frac{\partial \Phi\Lb Y - y_1,p_{1,T}; Y - y_2,p_{2,T}; q_T \Rb}{\partial \,\Lb Y - y_2\Rb} \,\,&=&\,\, \bas \int \frac{d^2 k_T}{2 \pi} K\Lb p_{2,T}, k_T; q_T\Rb \,\Phi\Lb Y - y_1,p_{1,T}; Y - y_2, k_T; q_T\Rb\,\nn\\
&+&\,\,\bas \,\, \phi_{\rm pr}\Lb Y - y_2, p_{2,T}\Rb\, \phi\Lb y_2  - y_1, \vec{p}_{2,T},\vec{p}_{1,T},q_T\Rb\vartheta\Lb y_2 - y_1\Rb;\label{EQBFKL22}
 \eea 
  where $\vartheta\Lb y_{12}\Rb$ is the step function. 
Comparing \eq{EQBFKL21} and \eq{EQBFKL22} with \eq{EQBFKL1}
 and \eq{EQBFKL2}, we see that we have found the exact form of the 
function
 $\Gamma\Lb \vec{p}_{1T},\vec{p}_{2T}; y_1, y_2\Rb $  for $\bas
 | y_1 - y_2| \,\gg\,1$. 
 
 Recall, that $\phi\Lb y_2  - y_1, \vec{p}_{2,T},\vec{p}_{1,T},q_T\Rb$ 
denotes
 the multiplicity of  gluons with rapidities $y_1$ and transverse momenta
 $p_{1,T}$ in the gluon with rapidity $y_2$ and transverse momentum $p_{2,T}$.

  These equations are written for $y_2\, \gg \,y_1\, \gg 1$. 
For $y_2 \sim y_1$ we can replace $ \phi\Lb y_2  - y_1, 
\vec{p}_{2,T},\vec{p}_{1,T},q_T\Rb$ by the DGLAP single
 parton density. However,  
  we  discuss the Bose-Einstein correlation
 which are essential at $y_1 = y_2$. In the kinematic
 region $\bas |y_2 - y_1| \ll 1$ we  need to re-write
 the non-homogeneous term.


\begin{boldmath}
\subsection{Equations for $\bas |y_2  - y_1| \,\ll\,\,1$.}
\end{boldmath}
  
 In the framework of the LLA,  the gluon with the fraction of 
energy $x$ can produce two gluons with $x_1 \approx x_2 \, \ll\,x$ 
  in the subsequent decay  $g(x ,p_T) \to g(x', p'_T) +g(x_1,p_{1,T})
 $ and  $g(x' ,p'_T) \to g(x'',p''_T) +g(x_2,p_{2,T}) $ as  is shown
 in \fig{bfkl3p}-b.  We calculate the contributions of these decays to
 the partonic wave function using  light-cone perturbative
 theory (see Refs. \cite{KOLEB,GBS,Cruz-Santiago:2015dla}). The
 wave function of \fig{bfkl3p}-b can be written in the following form
 \bea \label{WF2G}
 \Psi\Lb\{x_i,p_{i,T}\},x'',p''_T;x_1,p_{1,T};x_2,p_{2,T}\Rb\,&=&\,
 \Gamma^{\sigma,\delta}_\beta \Lb p' \to p'' + p_2\Rb \epsilon^{*\lambda_2}_\delta(p_2)  \epsilon^{\lambda''}{\sigma}(p'') \theta(p''^+) \theta(p^+_2)\frac{1}{D_{n+2}}\\
 &\times&\Gamma^{\alpha, \beta,\gamma}  \Lb p\to p' + p_1\Rb\epsilon^\lambda_\alpha(p)\epsilon^{*\lambda_1}_\beta(p_1)\theta(p'^+)\theta(p_1^+)\frac{1}{D_{n+1}}\frac{1}{p^+}\Psi\Lb\{x_i,p_{i,T}\},; x, p_T\Rb\nn
  \eea
   where  polarization vectors $\epsilon^\lambda_\mu(p)$ are  defined as
   \beq \label{POLVEC}
   \epsilon^\lambda_\mu(p) \,\,=\,\,\Bigg( 0, \frac{2\,\vec{\epsilon}^\lambda_\perp \cdot \vec{p}_T}{\eta \cdot p}, \vec{\epsilon}^\lambda_\perp\Bigg);~~~~~\eta\,=\,(0,1,0,0) ~~~~~~\mbox{and} ~~~~ \vec{\epsilon}^{\pm}_\perp\,\,=\,\,\frac{1}{\sqrt{2}}( \pm 1, i)
   \eeq
 The light-cone denominators are defined as
 \beq \label{D}
 D_{n+1}\,=\,p^-_1 + p'^- + \sum_{i=1}^{n-1} p^-_i \,-\,P^- \, \xrightarrow{x_1 \ll x}\, \frac{p^2_{1,T}}{ x_1 P^+} ;~~D_{n+2} \,=\,\,p^-_1 + p^-_2 + p''^{-} +\sum_{i=1}^{n-1} p^-_i \,-\,P^-\xrightarrow{x_1 =x_2\ll x}\,\frac{p^2_{1,T}}{ x_1 P^+}\,  + \, \frac{p^2_{2,T}}{ x_2 P^+};
 \eeq
 where $P^- $ is the light-cone energy of the incoming hadron. In
 \eq{WF2G} we have omitted the color indices. 
 
 $\Gamma^{\alpha, \beta,\gamma}\Lb k_3 \to k_2   + k_1\Rb $ is the
 triple gluon vertex for the decay $g(x_3,k_3) \,\to\,g(x_1,k_1) \,+\,g(x_2,k_2)$    which takes the following form  ( see Table 2 of Ref.  \cite{Cruz-Santiago:2015dla}):
{\small  \beq \label{V+-}
\Gamma^{+\to+ +}= 2 i g x_3 v^*_{21};~~\Gamma^{+\to+ -}= 2 i g x_1 v^*_{32};~~\Gamma^{+\to - +}= 2 i g x_2 v^*_{13}; ~~\mbox{with}~~v_{ij} =\vec{\epsilon}^+_\perp \cdot\Big(\frac{\vec{k}_{i,\perp}}{x_i} - \frac{\vec{k}_{i,\perp}}{x_i}\Big);
\,\,\,  v^*_{ij} =\vec{\epsilon}^-_\perp \cdot\Big(\frac{\vec{k}_{i,\perp}}{x_i} - \frac{\vec{k}_{i,\perp}}{x_i}\Big);
  \eeq}
  
  Plugging \eq{WF2G} into \eq{ISP} and summing over all polarizations and
 colors, we obtain 
  \beq \label{SPEQY}
 \Phi_{\rm sp}\Lb Y - y_1,p_{1,T}; Y - y_2,p_{2,T}; q_T\Rb\,\,=\,\,\bas^2\,\int^Y_{y_2} d y'  
\,V\Lb \vec{p}_{1,T}, \vec{p}_{2,T},\vec{q}_T\Rb  \phi\Lb Y - y', \vec{p}_{1,T} +  \vec{p}_{2,T} \Rb
  \eeq

  where
    \beq \label{VV}
 V\Lb \vec{p}_{1,T}, \vec{p}_{2,T},\vec{q}_T\Rb\,\,=\,\,
\frac{\Lb \vec{p}_{1,T} + \h \vec{q}_T\Rb \cdot \Lb \vec{p}_{1,T} - \h \vec{q}_T\Rb}{\Lb \vec{p}_{1,T} + \h \vec{q}_T\Rb^2\,\Lb \vec{p}_{1,T} - \h \vec{q}_T\Rb^2}\,\,\frac{\Lb \vec{p}_{2,T} + \h \vec{q}_T\Rb \cdot \Lb \vec{p}_{2,T} - \h \vec{q}_T\Rb}{\Lb \vec{p}_{2,T} + \h \vec{q}_T\Rb^2\,\Lb \vec{p}_{2,T} - \h \vec{q}_T\Rb^2} 
  \eeq
 It should be noted that we obtain       \eq{VV}  by adding the 
diagram
 with a  different order of emission for gluons with $(x_1,p_{1,T})$
 and $(x_2, p_{2,T})$.
.
 
 Bearing   in mind, that   \eq{SPEQY} generates the non-homogeneous
 term for $y_2 = y_1$, we note that \eq{EQBFKL1} and \eq{EQBFKL2}
 can be re-written as one equation which has the following form:
\bea \label{EQBFKL3}
&&\frac{\partial \Phi\Lb Y - y,p_{1,T}; Y - y,p_{2,T}; q_T\Rb}{\partial \,\Lb Y - y\Rb} \,\,=\\
&&\,\,~~~~~~~~~\bas \int \frac{d^2 k_T}{2 \pi} \Bigg\{K\Lb p_{1,T}, k_T; q_T\Rb \,\Phi\Lb Y - y,k_{T}; Y - y, p_{2,T}; q_T\Rb
\,+\,K\Lb p_{2,T}, k_T; q_T\Rb \,\Phi\Lb Y - y, p_{1,T}; Y - y, k_T; q_T\Rb\Bigg\}\,\nn\\
&&
\,~~~~~~~~~~~~~~~~~~~~~~~~+\,\,\bas^2\,V\Lb \vec{p}_{1,T}, \vec{p}_{2,T},\vec{q}_T\Rb
\,\,\phi\Lb Y - y, \vec{p}_{1,T} + \vec{p}_{2,T}\Rb\nn
\nn
\eea
  

\vspace{0.5cm}

\,\,

\begin{boldmath}
\subsection{Solution  in the region of  $\bas |y_2  - y_1| \,\ll\,\,1$.}
\end{boldmath}
  
\vspace{0.5cm}

We solve \eq{EQBFKL3} at $q_T=0$,  considering its  
  Mellin transform:
\beq \label{MLN}
\Phi\Lb \omega, \gamma_1, \gamma_2\Rb\,\,=\,\int^\infty_0 d (  Y - y) \, e^{\omega (Y - y)}\,\int \frac{d^2 p_{1,T}}{ 2 \pi}\,\frac{d^2 p_{2,T}}{2 \pi} \Lb p^2_{1,T}\Rb^{- \gamma_1} \,\Lb p^2_{2,T}\Rb^{- \gamma_2}\,\Phi\Lb Y - y,p_{1,T}; Y - y,p_{2,T}; q_T = 0\Rb
\eeq

For $
\Phi\Lb \omega, \gamma_1, \gamma_2\Rb$ \eq{EQBFKL3}  can be re-written
 in the form:
\beq 
\label{EQMLN}
\omega\,\Phi\Lb \omega, \gamma_1, \gamma_2\Rb\,\,=\,\,\bas\,\Lb \chi\Lb\gamma_1\Rb\,+\,\chi\Lb \gamma_2\Rb\Rb\,\Phi\Lb \omega, \gamma_1, \gamma_2\Rb\,\,+\,\,H\Lb \gamma_1, \gamma_2\Rb \phi\Lb \gamma_1 + \gamma_2\Rb
\eeq
where\cite{BFKL,KOLEB} 
\beq \label{CHI}
\chi\Lb \gamma\Rb\,\,=\,\,2 \psi\Lb 1\Rb \,-\,\psi\Lb \gamma\Rb\,-\,\psi\Lb 1 - \gamma\Rb\,\,\xrightarrow{\gamma \to \h}\,\,\omega_0\,\,+\,\,D\,\Lb \gamma - \h\Rb^2\,+\,{\cal O}\Lb \Lb\gamma - \h\Rb^4\Rb\,\,\leftarrow \,\,\mbox{diffusion approximation}
\eeq
where $\psi(z) = d \ln \Gamma(z)/d z$ is the Euler psi-function (
 see formula {\bf 8.36} of Ref.\cite{RY}) and $\omega_0 \,=\,4 \ln 2 $
 and $D = 14\zeta(3)$, where $z(x)$ is Riemann zeta function ( see 
formulae
 {\bf 9.51 - 9.53} of Ref.\cite{RY}). 

Function $H\Lb \gamma_1, \gamma_2\Rb$ is equal to
{\small \beq \label{H}
H\Lb \gamma_1, \gamma_2\Rb\,\delta\Lb \gamma  - \gamma_1 - \gamma_2\Rb\,=\,\int \frac{d^2 p_{1,T}}{2 \pi}\,\frac{d^2 p_{2,T}}{ 2 \pi} \Lb p^2_{1,T}\Rb^{ - \gamma_1} \,\Lb p^2_{2,T}\Rb^{ - \gamma_2}\, V\Lb \vec{p}_{1,T}, \vec{p}_{2,T},\vec{q}_T = 0 \Rb \Lb \Lb \vec{p}_{1,T} + \vec{p}_{2,T}\Rb^2\Rb^{\gamma - 1}
\eeq}
\eq{H} can be re-written  after taking the integrals over the angle and 
$p_{1,T}$ as
\bea \label{H1}
&&H\Lb \gamma_1, \gamma_2\Rb\,\,=\,\,\,\int^1_0 d x \Lb x^{2 \gamma_1 -1}\,+\,x^{2 \gamma_2 - 1}\Rb \\
&&\times\,\,\Bigg( (1 - x)^{ 2(\gamma_1 + \gamma_2)}{}_2F_1\Lb \h, - \gamma_1 - \gamma_2,1,- \frac{4 x}{(1 - x)^2}\Rb \,+\,(1 + x)^{ 2(\gamma_1 + \gamma_2)}{}_2F_1\Lb \h, - \gamma_1 - \gamma_2,1,\frac{4 x}{(1 + x)^2}\Rb\Bigg)\nn
\eea

\begin{figure}[ht]
    \centering
  \leavevmode
      \includegraphics[width=7cm]{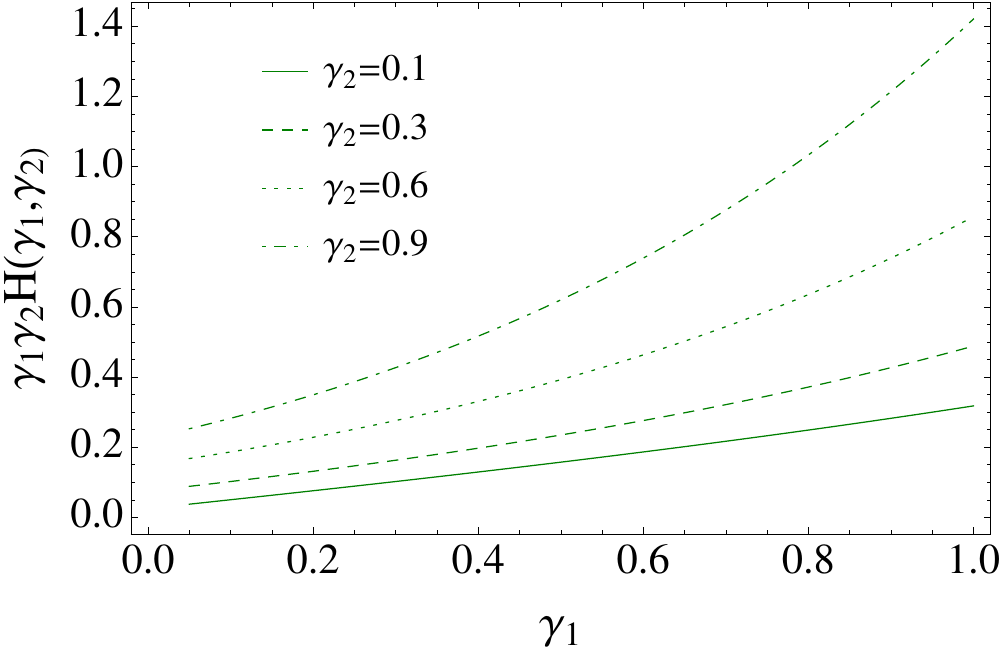}     
    \caption{The behavior of function $H\Lb \gamma_1, \gamma_2\Rb$.}
\label{h}
  \end{figure}

We illustrate the behavior of this function in \fig{h}. Note, that
 $H\Lb \gamma_1,\gamma_2\Rb$ is symmetric with respect to the change
 $\gamma_1 \to \gamma_2$ ( $H\Lb\gamma_1,\gamma_2\Rb = 
H\Lb\gamma_2,\gamma_1\Rb$).   

The particular solution to \eq{EQMLN} has a simple form:
\beq \label{SOL1}
\Phi\Lb \omega, \gamma_1, \gamma_2\Rb\,\,=\,\,\frac{H\Lb \gamma_1,\gamma_2\Rb \,\phi\Lb \omega, \gamma_1+\gamma_2\Rb}{\omega \,-\,\bas\Lb \chi\Lb \gamma_1\Rb\,+\,\chi\Lb \gamma_2\Rb\Rb}\,=\,\frac{H\Lb \gamma_1,\gamma_2\Rb \,\phi_{\rm in}\Lb \gamma_1 + \gamma_2\Rb}{\Lb \omega \,-\,\bas\Lb \chi\Lb \gamma_1\Rb\,+\,\chi\Lb \gamma_2\Rb\Rb\Rb \,\Lb \omega \,-\,\bas \chi\Lb \gamma_1 + \gamma_2\Rb \Rb}
\eeq
where$\phi_{\rm in}$ can be found from the initial conditions for the
 single gluon density.

From \eq{SOL1} we obtain the particular solution  for
 $\Phi\Lb Y - y,p_{1,T}; Y - y,p_{2,T}; q_T\Rb$
\bea \label{SOL2}
&&\Phi_{\rm part}\Lb Y - y,p_{1,T}; Y - y,p_{2,T}; q_T\Rb\,\,=\\
&&\!\!\!\int^{\epsilon + i \infty}_{\epsilon - i\infty}\frac{d \gamma_1}{2 \pi i}\int^{\epsilon + i \infty}_{\epsilon - i\infty}\frac{d \gamma_2}{2 \pi i}\,e^{ \gamma_1 \xi_1 \,+\,\gamma_2 \xi_2} \frac{\bas^2\,H\Lb \gamma_1,\gamma_2\Rb \,\phi_{\rm in}\Lb \gamma_1 + \gamma_2\Rb}{\bas\Lb \chi\Lb \gamma_1\Rb\,+\,\chi\Lb \gamma_2\Rb\,-\,\chi\Lb \gamma_1 + \gamma_2\Rb\Rb}\Bigg\{ e^{ \bas\Lb \chi\Lb \gamma_1\Rb\,+\,\chi\Lb \gamma_2\Rb\Rb\,\Lb Y - y\Rb}
\,\,-\,\,e^{ \bas  \chi\Lb \gamma_1 + \gamma_2\Rb\,\Lb Y - y\Rb}\Bigg\}\nn
\eea
where $\xi_1\,=\,\ln\Lb p^2_{1,T}\Rb$   and    $\xi_2\,=\,\ln\Lb p^2_{2,T}\Rb$.
The general solution will be a sum of the particular solution and the
 solution to the homogeneous equation, and it has the following form
\bea \label{SOL3}
&&\Phi\Lb Y - y,p_{1,T}; Y - y,p_{2,T}; q_T\Rb\,\,=\\
&&\,\,\Phi_{\rm part}\Lb Y - y,p_{1,T}; Y - y,p_{2,T}; q_T\Rb\,+\,
\!\!\!\int^{\epsilon + i \infty}_{\epsilon - i\infty}\frac{d \gamma_1}{2 \pi i}\int^{\epsilon + i \infty}_{\epsilon - i\infty}\frac{d \gamma_2}{2 \pi i}\,e^{ \gamma_1 \xi_1 \,+\,\gamma_2 \xi_2} \Phi_{\rm in}\Lb \gamma_1, \gamma_2\Rb\,e^{ \bas\Lb \chi\Lb \gamma_1\Rb\,+\,\chi\Lb \gamma_2\Rb\Rb\,\Lb Y - y\Rb}\,\,=\nn\\
&&\,\,\int^{\epsilon + i \infty}_{\epsilon - i\infty}\frac{d \gamma_1}{2 \pi i}\int^{\epsilon + i \infty}_{\epsilon - i\infty}\frac{d \gamma_2}{2 \pi i}\,e^{ \gamma_1 \xi_1 \,+\,\gamma_2 \xi_2} \Bigg\{ \Bigg[\frac{\bas H\Lb \gamma_1,\gamma_2\Rb \,\phi_{\rm in}\Lb \gamma_1 + \gamma_2\Rb}{\Lb \chi\Lb \gamma_1\Rb\,+\,\chi\Lb \gamma_2\Rb\,-\,\chi\Lb \gamma_1 + \gamma_2\Rb\Rb}\,+\,
\Phi_{\rm in}\Lb \gamma_1, \gamma_2\Rb\Bigg] e^{ \bas\Lb \chi\Lb \gamma_1\Rb\,+\,\chi\Lb \gamma_2\Rb\Rb\,\Lb Y - y\Rb}\nn\\
&&
\,\,-\,\,\frac{\bas H\Lb \gamma_1,\gamma_2\Rb \,\phi_{\rm in}\Lb \gamma_1 + \gamma_2\Rb}{\Lb \chi\Lb \gamma_1\Rb\,+\,\chi\Lb \gamma_2\Rb\,-\,\chi\Lb \gamma_1 + \gamma_2\Rb\Rb}e^{ \bas  \chi\Lb \gamma_1 + \gamma_2\Rb\,\Lb Y - y\Rb}\Bigg\}\nn\eea

The integrals over $\gamma_1$ and $\gamma_2$ in the first
 term of \eq{SOL3} can be  evaluated using the method of steepest
 descend with the saddle point for both $\gamma$'s close to 
$\h$, where we can use the diffusion approximation
 (see \eq{CHI}) for $\chi(\gamma)$. The values of
  $\gamma$'s at the saddle point are the following:
\beq \label{SPGA}
\gamma^{\rm SP}_1\,\,=\,\,\h - \frac{\xi_1}{2 \bas D \Lb Y - y\Rb};
~~~~~\gamma^{\rm SP}_2\,\,=\,\,\h - \frac{\xi_2}{2 \bas D \Lb Y - y\Rb};
\eeq
After integration over $\gamma_1$ and $\gamma_2$ in the vicinities of 
these saddle points we obtain the contribution:
\beq \label{SOL4}
\Phi\Lb Y - y,p_{1,T}; Y - y,p_{2,T}; q_T\Rb\,\,=\,\,\frac{1}{4 \pi}\frac{1}{\bas D \Lb Y - y\Rb}\Bigg[ \dots\Bigg]_{\gamma_1 = \gamma^{\rm SP}_1; \gamma_2 = \gamma^{\rm SP}_2}  \!\!\!\!\!\!\!\!\exp\Lb 2 \bas \omega_0\Lb Y - y\Rb  \,-\,\frac{\xi^2_1 + \xi_2^2}{4 \bas D \Lb Y - y\Rb}\Rb
\eeq
 This contribution is proportional to
 $\phi\Lb Y - y, \xi_1\Rb\,\phi\Lb Y - y, \xi_2\Rb$ and
 in  agreement with \eq{I2}.

In the second term the integration over $\gamma_1 + \gamma_2$ can 
be taken using the method of steepest descend, leading to 
\beq \label{SP2T}
(\gamma_1 + \gamma_2)^{\rm SP} \,=\,\h\,-\,\frac{\xi_1 + \xi_2}{2
 \bas D \Lb Y - y\Rb}
\eeq

This integration generates the contribution which is proportional
 to $\exp\Lb \bas \omega_0 \Lb Y - y\Rb \,-\,\Lb \xi_1 + \xi_2\Rb^2/\Lb 4
 \bas D \Lb Y - y\Rb\Rb\Rb$. Comparing this contribution with \eq{SOL4}, 
one can see that it is suppressed at large values of ( $Y - y$) .

~

~

  \section{The interference diagram in the BFKL evolution}
\begin{figure}[ht]
    \centering
  \leavevmode
  \begin{tabular}{c c c c c}
      \includegraphics[width=5.6cm]{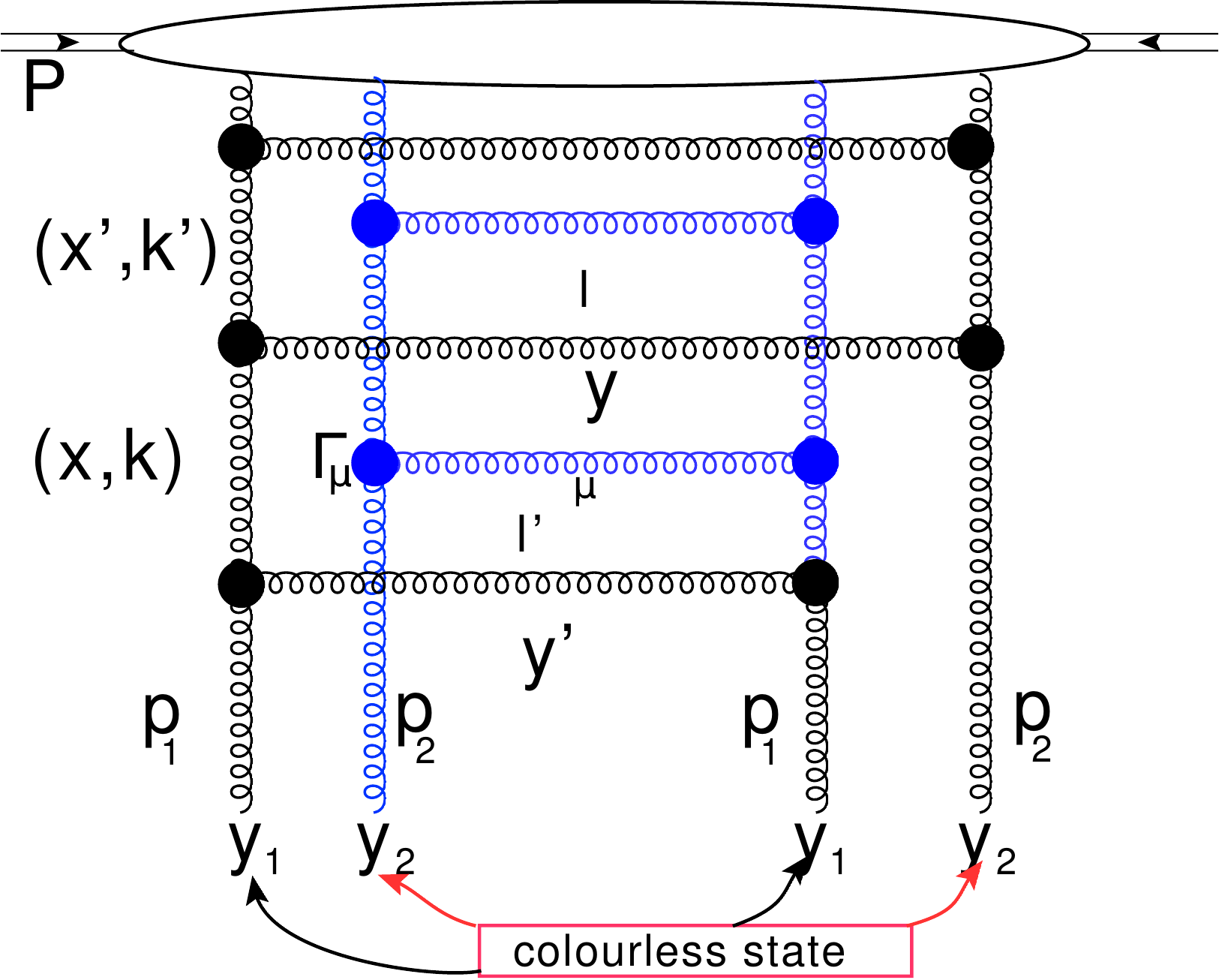}  &~~~& \includegraphics[width=4.5cm]{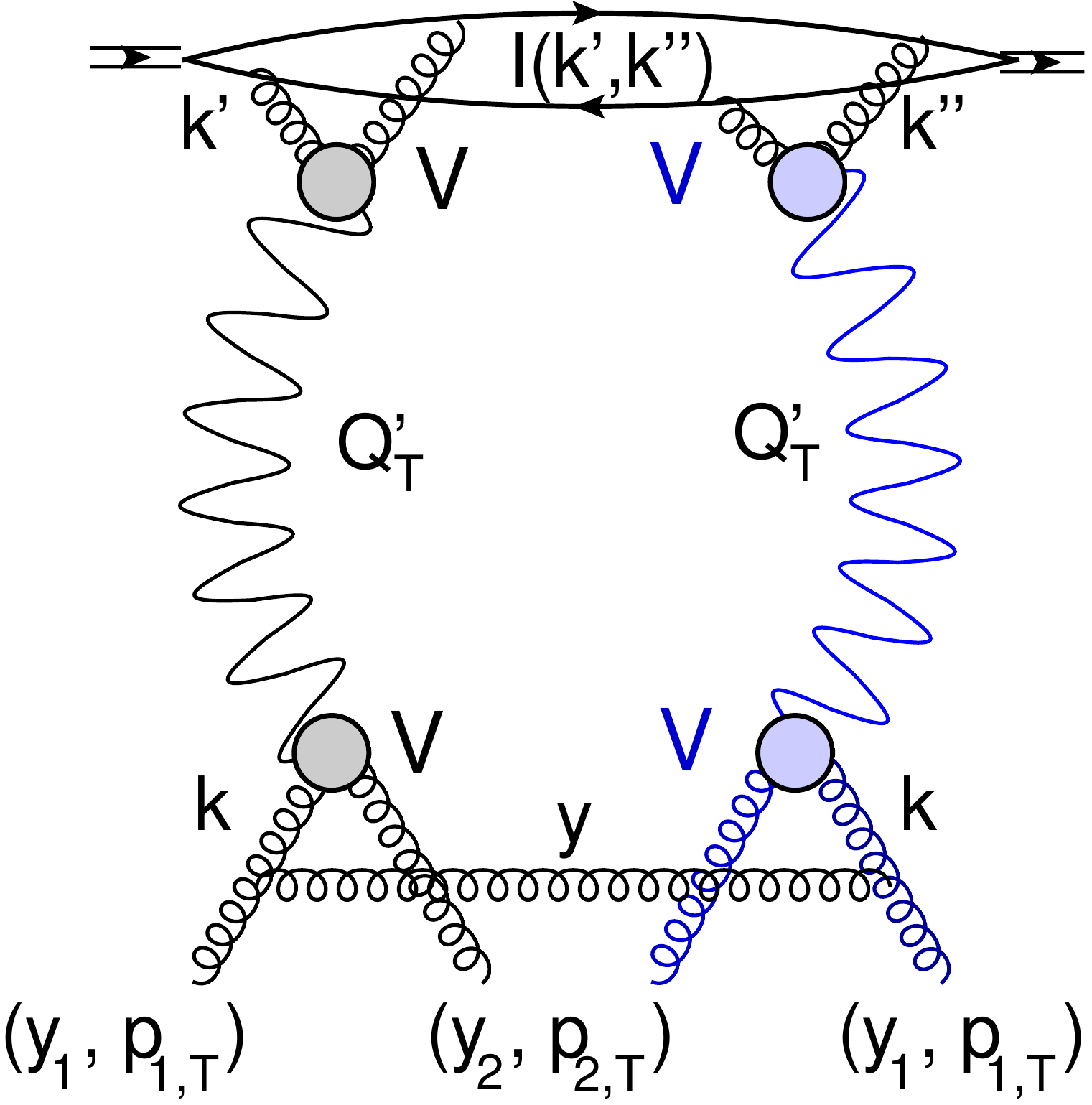}& ~~~& \includegraphics[width=6cm,height=4.5cm]{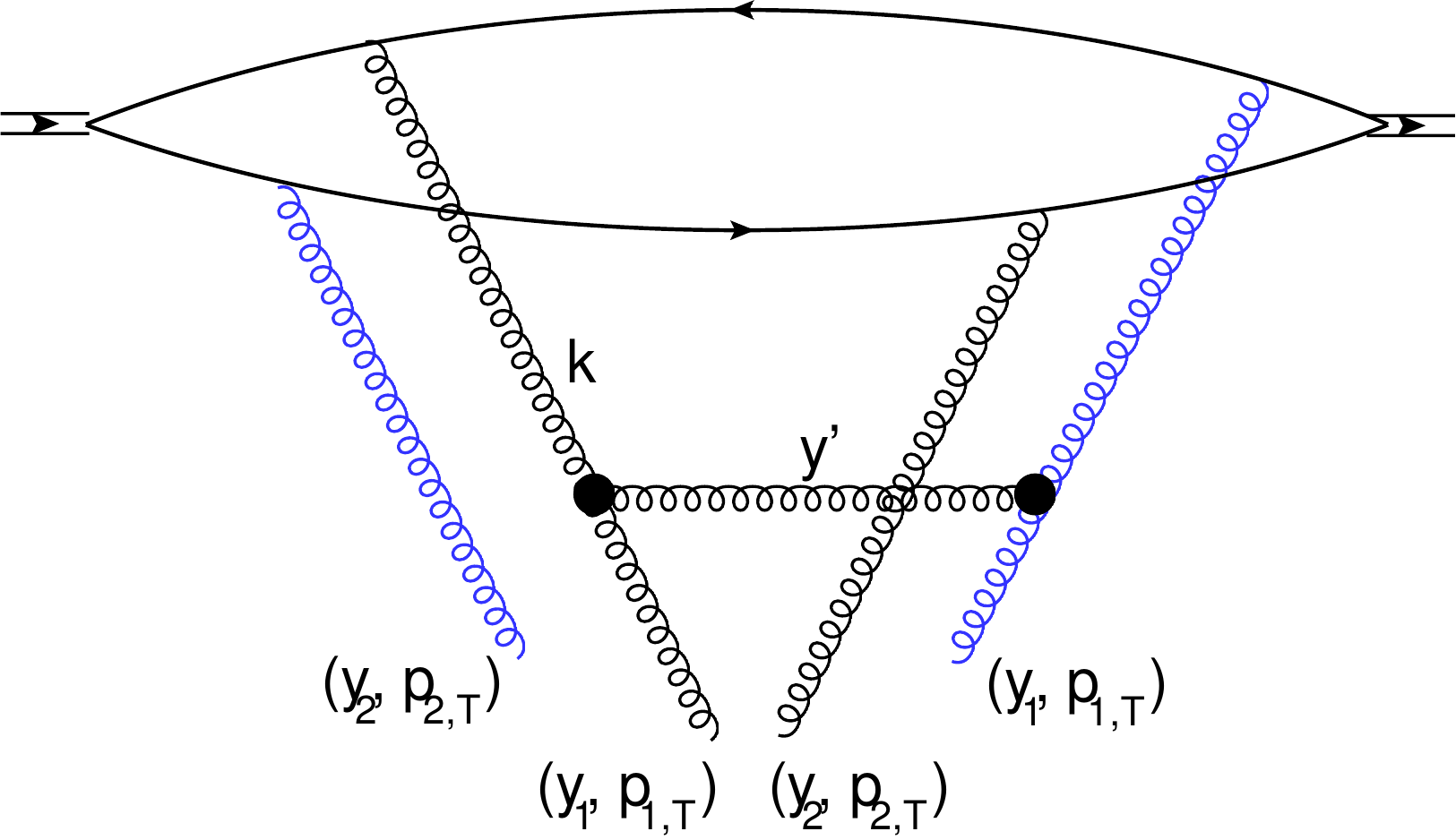} \\
      \fig{intdi}-a &  & \fig{intdi}-b& & \fig{intdi}-c\\
      \end{tabular}
         \caption{The interference diagram.\fig{intdi}-a:  For gluons with
 rapidities larger
 than $y'$, the $t$-channel gluons, which are shown  by the same 
color
 helical lines,   are in a colorless state in the $t$-channel. The gluons
 with $y_1$ and with $y_2$   are in a
colorless
 states as  indicated   by arrows in \fig{I2}. The black
 circles denote the Lipatov vertex $\Gamma_\mu$. \fig{intdi}-b:
 This is  the diagram of \fig{I2}-a, which is re-drawn through the
 BFKL Pomerons  denoted by wavy lines. Their colors in the figure
 indicate the gluons that constitute  the pomerons
 (see \fig{intdi}-b). The gluons with $y_1,p_{1,T}$ as well as gluons
 with $y_2, p_{2,T}$ are in colorless states. $ \vec{Q}'_T \,=\,\vec{k}
 - \vec{p}_{2,T}$.  $V$  denote the vertices of gluon-Pomeron interaction. 
\fig{intdi}-c: This diagram is the Born approximation of the diagram of 
\fig{intdi}-b in the case of an onium target. This diagram is in accord 
with
 \eq{I2}.}
\label{intdi}
  \end{figure}

    The interference diagram is shown in \fig{intdi}-a. In this diagram
 the $t$-channel gluons with the same color and with rapidities larger 
than
 $y'$, are in  colorless states. For rapidities, that are less than
 $y'$, $t$ channel gluons with rapidities $y_1$ and with $y_2$ are in
 a colorless state. The arguments for such a color structure of this
 diagram stem from the first diagram with the exchange of  two identical
 gluons, shown in \fig{intdi}-b. In this diagram all emitted gluons with
  rapidities larger than $y'$,  can be absorbed in the solution of the
 evolution equation without the Bose-Einstein enhancement, and can be used
 as a solution of \eq{I2}.
 In this solution the double gluon density  can be viewed as the exchange
 of two BFKL Pomerons, shown in \fig{intdi}-b.   
    
    These Pomerons carry  transferred momenta $\vec{Q}'_T$ and $- 
\vec{Q}'_T$, respectively,  where $\vec{Q}'_T \,=\,\vec{k}\,-\,\vec{p}_{2,T}$.
 Therefore, we  first  need to  deal with  the BFKL Pomeron with 
non
 zero  transfer momentum .  However, before discussing this 
problem we
 calculate the diagram of \fig{intdi}-c, which is the diagram of 
\fig{intdi}-b in the Born approximation. Denoting the wave function
 of the colorless dipole ( the onium state of a  heavy quark and
 antiquark)  by 
  $  \Psi\Lb q_T, z\Rb$, where $q_T $ is the transverse momentum of 
the quark and $z$ its fraction of the energy. We obtain that the 
 component of the gluonic wave function with one emitted gluon
 with transverse momentum $p_{1,T}$ and rapidity $y_1$ is equal
 to \cite{MUDI,KOLEB}
  \beq \label{ID1}
  \Psi^{(1)}\Lb \vec{q}_T, z; \vec{p}_{1,T}, y_1\Rb\,=\,g \lambda^a
 \frac{\vec{p}_{1,T} \cdot \vec{\epsilon}^\lambda_1}{p^2_{1,T}}\Bigg( \Psi\Lb \vec{q}_T, z\Rb \,\,-\,\, \Psi\Lb \vec{q}_T + \vec{p}_{1,T}, z\Rb  \Bigg)
  \eeq
  where $\lambda^a$ denotes the Gell-Mann matrix and 
$\vec{\epsilon}^\lambda$
 is the polarization vector of the gluon with the helicity $\lambda$.
 The single gluon density has the form
  \beq \label{ID2}
  \phi\Lb y_1, p_{1,T}\Rb\,\,=\,\,\int^1_0 \frac{d z}{z (1-z)}\int  d^2 q_T | \Psi^{(1)}\Lb \vec{q}_T, z; \vec{p}_{1,T}, y_1\Rb |^2\,\,=\,\, \frac{\as C_F}{\pi}\frac{1}{p^2_{1,T}}\,\Lb G\Lb 0\Rb - G\Lb 4 p^2_T\Rb\Rb
  \eeq
  In the integral over $z$, the lower limit is $e^{-Y+ y}$, but we assumed
 that this integral is convergent  and   we can safely take  this limit
 equal to zero.
  \beq \label{ID3}
  G\Lb \vec{p}_T\Rb\,\,=\,\,\int \frac{d z}{z (1-z)}\int d^2 r e^{  i \vec{p}_T \cdot \vec{r}} | \Psi\Lb r, z\Rb|^2~~~~~ \mbox{where}~~~\int d^2 r \int \frac{d z}{z (1-z)}  | \Psi\Lb r, z\Rb|^2 \,\,=\,\,1;
  \eeq  
  
  The emission of the second gluon with $y_2$ and $p_{2,T}$ leads to
 the wave function
  \bea \label{ID4}
 && \Psi^{(1,1)}\Lb \vec{q}_T, z; \vec{p}_{1,T}, y_1;\vec{p}_{2,T}, y_2\Rb\,=\\
 &&\,g^2 \lambda^a \lambda^b\frac{\vec{p}_{1,T} \cdot \vec{\epsilon}^\lambda_1}{p^2_{1,T}}\frac{\vec{p}_{2,T} \cdot \vec{\epsilon}^\lambda_2}{p^2_{2,T}}
  \Bigg( \Psi\Lb \vec{q}_T, z\Rb \,\,-\,\, \Psi\Lb \vec{q}_T + \vec{p}_{1,T}, z\Rb   \,\,-\,\,\Psi\Lb \vec{q}_T + \vec{p}_{2,T}, z\Rb \,\,+\,\,\Psi\Lb \vec{q}_T + \vec{p}_{1,T} + \vec{p}_{2,T}, z\Rb\Bigg)\nn
  \eea

 Using  the expression for the Lipatov vertex $\Gamma_\mu$
 which  has the form \cite{BFKL,KOLEB}
             \beq \label{V}
 \Gamma_\mu\Lb k' , k\Rb\,\,=\,\,
2 g \,f^{abc}\Bigg(  k'_{T, \mu}\,-\,\frac{k'^2_{T}}{l^2_T} \,l_{T,\mu}\Bigg) 
 \eeq
 where $ l_\mu\,=\,k'_{T,\mu}\,-\,k_{T,\mu}$ is the momentum of the emitted
 gluon, as well as      
 \eq{ID4}  we obtain for \fig{intdi}-c the following contribution
   \bea \label{ID5}
   A\Lb \fig{intdi}-c\Rb\,\,&\propto&\,\,\int d^2 k_T\,I\Lb \vec{p}_{2,T}, \vec{k}_T\Rb\frac{\Lb \vec k_T \cdot \vec{p}_{2,T}\Rb^2}{\Lb p^2_{2,T} \,k^2_T\Rb^2}\underbrace{\frac{k^2_T\,p^2_{1,T}}{\Lb \vec{k}_T - \vec{p}_{1,T}\Rb^2}}_{\Gamma_\mu \Gamma_\mu}\frac{1}{p^4_{1,T}}\nn\\
   &\,\,=\,\,&\,\,\int d^2 k_T\,I\Lb \vec{p}_{2,T}, \vec{k}_T\Rb\frac{\Lb \vec k_T \cdot \vec{p}_{2,T}\Rb^2}{\Lb p^2_{2,T} \,k^2_T\Rb^2}\frac{k^2_T\,p^2_{1,T}}{\Lb \vec{k}_T\,-\,\vec{p}_{2,T} - \Delta\vec{p}_{12,T}\Rb^2}\frac{1}{p^4_{1,T}}\,   \eea  
  where $\Delta \vec{p}_{12,T} \,=\vec{p}_{1,T} - \vec{p}_{2,T}$ and \,\,  $ \Lb \vec{p}_{2,T}, \vec{k}_T\Rb$ is equal to (see Ref.\cite{BEC5})   
  \beq \label{ID6}
   I\Lb \vec{p}_{2,T},  \vec{k}_{T}\Rb = 
 2\,+\, G\Lb 2 (\vec{k}_T + \vec{p}_{2,T})\Rb +  G\Lb 2 (\vec{k}_T - \vec{p}_{2,T} )\Rb  - G\Lb 2    \vec{k}_T \Rb - G\Lb - 2 \vec{k}_T\Rb\, - \,G\Lb 2   \vec{p}_{2,T }\Rb\,-\,G\Lb - 2   \vec{p}_{2,T }\Rb;    
  \eeq 
 
   In \eq{ID4}  for simplicity, we  have omitted all color coefficients 
and coupling constants.
   
   We see that the largest contribution stems from the region: $|\vec{p}
_{2,T} - \vec{k}_T| \propto 1/r$, where $r$ is the size of the dipoles.
 Assuming that $p_{1,T}$ and $p_{2,T}$ are larger that $1/r$ this region
 leads to the contribution which is equal
   \beq \label{ID7}
  A\Lb \fig{intdi}-c\Rb\,\,\propto\,\,\frac{1}{p^2_{1,T}\,p^2_{2,T}} \int
 \frac{ d^2 k_T}{\Lb \vec{k}_T\,-\,\vec{p}_{2,T} -
   \Delta\vec{p}_{12,T}\Rb^2}\,   I\Lb \vec{p}_{2,T},  \vec{k}_{T}\Rb \,\sim \,\frac{1}{p^2_{1,T}\,p^2_{2,T}}\pi \int d q^2 \frac{1}{|q^2 - |\Delta\vec{p}_{12,T}|^2|}I(q)
  \eeq
  with $\vec{q}=\vec{k}_T - \vec{p}_{2,T}$. \eq{ID7} generates
 $|q|\sim 1/r \,\ll\,p_{i,T}$ and shows that the Bose-Einstein
 enhancement is essential for $|\Delta\vec{p}_{12,T}| \,\ll\,p_{i,T}$
 and the value of $|\Delta\vec{p}_{12,T}| $ is determined by
the scale of the initial conditions for the double gluon density.
    
   In other words we conclude that the first diagram leads to the double
gluon density, which can be written in the form
   \beq \label{ID8}
   \Phi^{\rm int}\,\propto\,\phi\Lb y_1,p_{1,T}\Rb\,\phi\Lb y_2,p_{2,T}\Rb \left\{\begin{array}{l}\,\propto\,\,{\rm Const}\,\,\,\,\,\mbox{for }\,\,\, |\Delta\vec{p}_{12,T}|\,r\,\,\leq \,1;\\ \\
\,\frac{1}{r^2\,|\Delta\vec{p}_{12,T}|^2} \,\,\,\,\,\mbox{for}\,\,\, |\Delta\vec{p}_{12,T}|\,r\,\,\geq \,1; \end{array}
\right.  \,\sim\,\,\frac{\phi\Lb y_1,p_{1,T}\Rb\,\phi\Lb y_2,p_{2,T}\Rb}{r^2\,|\Delta\vec{p}_{12,T}|^2\,\,+\,\,1}
 \eeq    
In other words, the Bose-Einstein enhancement  increases the double gluon
 densities for $\vec{p}_{1,T} \to \vec{p}_{2,T}$, as  is expected and
 has been demonstrated in the correlation functions\cite{BEC1,BEC2,BEC3,
BEC4,BEC5,BEC6}.

 Returning to  the diagram of \fig{intdi}-b we see
 that generally the integration over $k_T$ enters the integration
 over momentum transfer of the BFKL Pomeron:    $\vec{Q}'_T \,=\,
\vec{p}_{2,T} - \vec{k}_T$,  and integration over $k_T$ that characterize
 the size of the dipole in the Pomeron vertices (see \fig{intdi}-b).  The
 typical transverse momentum in the BFKL Pomeron vertices are about 
 $p_{1,T}(p_{2,T})$ or about that of the saturation scale, at the rapidity
 of the vertex. On the other hand, the typical $Q'_T\,\sim\,1/r$ where
 $r$ is the size of the largest dipole of the  two interacting dipoles,  
 which constitute  the exchange of
  the BFKL Pomeron. For the diagrams of \fig{intdi}-b this largest dipole
 has a size of the hadron, whose double parton density we discuss.

    The Green function of the  BFKL Pomeron  $ G\Lb \vec{r},\vec{R};
 \vec{Q}_T; Y\Rb$  is known in the mixed representation,  where $r$
 and $R$ are the sizes of two
 interacting dipoles, $Q_T$ denotes the momentum transferred by
 the Pomeron, and  $Y$  the rapidity between the two dipoles.
 This Green function  has the following form\cite{LIP,NAPE}:
  \beq \label{BFKLGF}
G\Lb \vec{r},\vec{R}; \vec{Q}_T; Y\Rb \,=\,\,\frac{1}{16}\sum^{\infty}_{n=
 - \infty}\int^{\infty}_{-\infty} d \nu\frac{1}{\Lb \nu^2 +  
 \qu(n-1)^2\Rb\,\Lb \nu^2 + \qu(n+1)\Rb} V_{\nu,n}\Lb \vec{r},  \vec{Q}_T\Rb\,V^*_{\nu,n}\Lb\vec{R},\vec{Q}_T\Rb\,e^{ \omega\Lb \nu,n\Rb \,Y}
  \eeq
where 
  \beq \label{OMEGA}
  \omega\Lb\nu,n\Rb \,\,=\,\,2\,\bas {\rm Re}\Lb \psi\Lb \h + 
\h |n| + \nu\Rb - \psi\Lb 1 \Rb \Rb; ~~~~~ \mbox{and}~~~
 \omega\Lb\nu,0\Rb \,\,=\,\,\bas \chi\Lb \h + i \nu\Rb;
 \eeq
 with $n = 0,1,3 \dots$ and 
 $\chi\Lb \gamma\Rb$ from \eq{CHI}. 
 
     Each term in \eq{BFKLGF} has a very simple structure, being the
 typical contribution of a Regge pole exchange: the product of two
 vertices, which depend on the size of the dipole and $Q_T$, and the
 Regge-pole propagator $e^{ \omega\Lb \nu,n\Rb \,Y}$. From \eq{OMEGA}
 one can see
 that at large $Y$ the main contribution comes from the term with
 $n = 0,$ and in what follows we will concentrate on this particular  
term. 

   The vertices with $n=0$ have been  determined in 
Refs.\cite{LIP,NAPE},
 and they have a simple form in the complex number representation
 for the point on the two dimensional plane: viz.
   \beq \label{CN}
\mbox{For } \vec{r}(x,y)  : \,\,   \rho = x + i y; \,\, \rho^*= x - i y;~~~~~
\mbox{For } \Qv(Q_x, Q_y)  : \,\,  q = Q_x + i Q_y;\,\, q^*= Q_x - i Q_y;
\eeq
  Using  this notation the vertices  
  have the following structure:
  \beq\label{V}
  V_{\nu}\Lb \rv,\Qv\Rb\,=\,r\,\Lb Q^2_T\Rb^{i \nu}\,\Gamma^2\Lb 1 - i \nu\Rb\Bigg\{
  J_{-i \nu}\Lb \qu q^* \rho\Rb \,J_{- i \nu}\Lb \qu q \rho^*\Rb  \,\,-\,\, J_{i \nu}\Lb \qu q^* \rho\Rb \,J_{ i \nu}\Lb \qu q \rho^*\Rb \Bigg\}
     \eeq 
     
         At $Q_T \to 0$ this vertex takes the form:
    \beq \label{VSQ}
V_{\nu}\Lb \rv,\Qv\Rb\,\xrightarrow{Q_T r\,\ll\,1}  \,\,r\,\Bigg\{
  \left(\frac{r^2}{2^6}\right)^{-i \nu }\, -\, \left(Q^2\right)^{i \nu } \left(\frac{Q^2 r^2}{2^6}\right)^{i \nu }\Bigg\}
   \eeq  
        Using  that 
     \beq \label{ASJ}
     J_{-i \nu}\Lb z \Rb\,\,\xrightarrow{z\,\gg\,1}\,\,\sin\Lb
 \qu \pi + z + \h  i \pi \nu\Rb \sqrt{\frac{2}{\pi}} \sqrt{\frac{1}{z}}
     \eeq
     at  $\nu \ll 1$ we obtain for $Q^2_T r^2 \gg 1$
      \beq \label{VLQ}
 V_{\nu}\Lb \rv,\Qv\Rb\,\,\xrightarrow{Q_T r\,\gg\,1} \,\, \Lb Q^2_T\Rb^{i \nu}\,
 \Gamma^2\Lb 1 - i \nu\Rb\,   \,\cos\Lb \h \Qv \cdot \rv\Rb\, 
 \frac{4 
 \,i\, \nu}{Q_T}
 \eeq
       
        Returning  to \eq{BFKLGF}, one can see that the exchange of 
the
 BFKL Pomeron turns out to be small for $R\,Q_T\,>\,1 $, where $R$ is the
 size of the
 larger of the  two interacting dipoles. In the diagram of \fig{intdi}-b,
 the size of the smallest dipole is about $r \propto 1/p_{2,T}$, while $R 
$
 is the dipole in the hadron which has a size of the order $1/\mu_{\rm 
soft}$, 
  $\mu_{\rm soft}$ denotes the soft scale. In other words, we expect 
that $Q_T$  of the BFKL Pomerons is rather small,  $Q_T \leq\,\mu_{\rm
 soft}$. Since $|\vec{Q} '_T| \,=\,|\vec{k}_T - \vec{p}_{2,T} | \leq \mu_{\rm
 soft} \,\ll\,p_{2,T}$ we safely use for vertex $V_{\nu}\Lb \rv,
\Qv\Rb$ \eq{VSQ}     which gives for the vertex in the momentum representation:
       \beq \label{VMR}
     V_{\nu}\Lb \vec{k}_T,\Qv\Rb\,\, =\,\,r^2\,\int d^2 r e^{ - i \vec{r} \cdot \vec{k}_T }\,V_{\nu}\Lb \rv,\Qv\Rb
     \eeq
 the following expression:
 \beq \label{VMR1}
  V_{\nu}\Lb \vec{k}_T,\Qv\Rb \,\,=\,\, 2 \pi \Bigg( \frac{\Gamma\Lb \h + i \nu\Rb}{\Gamma\Lb \h  - i \nu\Rb}\Lb 2\,k^2_T\Rb^{ - \h + i \nu} \,\,-\,\,\frac{\Gamma\Lb \h - i \nu\Rb}{\Gamma\Lb \h  + i \nu\Rb}\Lb Q^2_T  \Rb^{2 i \nu}   \Lb 2 k^2_T\Rb^{-\h + i \nu}\Bigg)
  \eeq
  Actually, the second term does not contribute to the scattering amplitude at
 small $Q_T$ (see Refs.\cite{LIP,GOLELY}), and therefore the diagram of 
\fig{intdi}-b gives the following contribution:
  \bea 
  \frac{\partial \Phi^{\rm int}\Lb  Y - y, p_{1,T}, p_{2,T}\Rb}{\partial (Y - y) } \,&=&\,\frac{\bas}{N^2_c - 1}\,\int^{-i \epsilon + \infty}_{-i \epsilon -  \infty} \frac{ d \nu_1}{2 \pi} \int^{-i \epsilon + \infty}_{-i \epsilon -  \infty} \frac{ d \nu_2}{2 \pi}  
  e^{ \bas \Lb  \chi\Lb \nu_1\Rb +  \chi\Lb \nu_2\Rb\Rb \Lb Y - y\Rb}\label{VMR21}\\
  &\times& \int \frac{d^2 k'_T d^2 k''_T}{( 2 \pi)^4} \,I\Lb \kv'_T,\kv''_T\Rb  V_{\nu_1}\Lb \kv'_T, \vec{Q}'_T\Rb\, V_{\nu_2}\Lb \kv''_T, \vec{Q}'_T\Rb\,
  \int \frac{d^2  Q'_T}{\Lb \Delta\vec{p}_{12,T} - \vec{Q}'_T    \Rb^2}     \Lb 2\,k^2_T\Rb^{ - 1 + i \nu_1 + i \nu_2}   \nn\\
  &=&  \,\frac{\bas}{N^2_c - 1}  \int \frac{d^2  Q'_T}{\Lb
 \Delta\vec{p}_{12,T} - \vec{Q}'_T    \Rb^2} \,\Phi_{ 2
 \pom}\Lb Y - y, p_{2,T}, p_{2,T}; Q'_T\Rb \label{VMR2}
 \eea
  where $\Phi_{2 \pom}$  denotes the double gluon density due to 
exchange of
 two BFKL Pomerons. All other notations are  shown in
\fig{intdi}-b. 
The second line of the equation is written assuming that $\nu_1\, \ll\,
 1$ and $\nu_2\,\ll\,1$ at high energies, in accord with the diffusion
 approximation (see \eq{CHI}).  It is easy to see that this contribution
 is the Fourier  transform of the emission term with $\rho^{(2)}$ in 
\eq{EVRO2NC} 
in momentum representation. We need to add the gluon reggeization term to
 \eq{VMR21} which is the Fourier transform of the second term with 
$\rho^{(2)}$
 in \eq{EVRO2NC}. Therefore, we do not see any other contribution except the
 Bose-Einstein enhancement in the first diagrams.
  In the last line of the equation we consider $ Q'_T \,\ll\,p_{2T}$ and 
replace $k_{T} $ by $p_{2,T}$. From \eq{VMR2} one can see that 
the double
 parton density with $q_T =0$ (see
  \eq{I1}) can be obtained only if we know the double parton density for
 $q_T \neq 0$.
  For $q_T \neq 0$ the diagram of \fig{intdi}-b can be re-written in the form:
 \beq \label{VMR3}
   \frac{\partial \Phi^{\rm int}\Lb  Y - y, p_{1,T}, p_{2,T}; q_T \Rb}{\partial (Y - y) } \,=  
  \frac{\bas}{N^2_c - 1}  \int \frac{d^2  Q'_T}{\Lb \Delta\vec{p}_{12,T} - \vec{Q}'_T -\vec{q}_T   \Rb^2} \,\Phi_{2 \pom}\Lb Y - y, p_{2,T}, p_{2,T}; \vec{Q}'_T\Rb 
  \eeq    
  
  \section{BFKL evolution with  Bose-Einstein enhancement}

     We need to change \eq{EQBFKL3} by adding the interference diagram.
 To do this  we have to change \eq{VMR21}, replacing  $\Phi_{2 \pom}\Lb Y
 - y, k_{T}, p_{2,T}; Q'_T\Rb $ by $  \Phi\Lb Y - y, k_{T}, p_{2,T}; Q'_T\Rb$
 and taking into account the complete BFKL kernel of \eq{KER}. Therefore, the
 contribution of the interference diagram to the evolution equation takes the
 form
    \beq \label{BFKLBE1}
   \frac{\partial \Phi^{\rm int}\Lb  Y - y, p_{1,T}, p_{2,T}; q_T \Rb}{\partial (Y - y) }\,=\,\int d^2 k_T\,K\Lb \vec{p}_{1T}-\vec{q}_T,\kv_{ T}; \vec{Q}'_T\Rb  \Phi\Lb Y - y, k_{T}, p_{2,T}; Q'_T\Rb
   \eeq
   where $K$ denotes the kernel of \eq{KER}.  In \eq{BFKLBE1} we  
have taken into
 account that $q_T \neq 0$.  Substituting this kernel we  reduce
 \eq{BFKLBE1} to the form
   \bea \label{BFKLBE1}
   &&\frac{\partial \Phi^{\rm int}\Lb  Y - y, p_{1,T}, p_{2,T}; q_T \Rb}{\partial (Y - y) }\,=\\
  && \, \frac{2\,\bas}{N^2_c - 1} \Bigg\{\int
 d^2 k_T\underbrace{\frac{1}{\Lb \vec{k}_T -
 \vec{p}_{1,T} - q_T\Rb^2}}_{\mbox{emission kernel}}
  \Phi\Lb Y - y, k_{T}, p_{2,T}; Q'_T\Rb\,\,-\,
\,\underbrace{\omega_G\Lb\vec{p}_{2,T} - \vec{q}_T\Rb}_{\mbox{reggeization
 kernel}}\,\Phi\Lb Y - y, p_{2,T}, p_{2,T}; q_T\Rb\Bigg\} \nn  \eea

   As we have discussed in the previous section, the typical $Q'_T$
 is determined by the soft scale from the initial condition, as in
 \fig{intdi}-b,  or by the saturation scale at rapidity $y' > y$.
 Both  are much smaller than $p_{iT}$, or the saturation momentum
 at rapidity $y$. Hence, we can neglect $\vec{Q}'_T = \vec{k}_T -
 \vec{p}_{1T}$, as it is  much smaller than $p_{i,T}$. 
   Finally, \eq{BFKLBE1} takes the form:
      \bea \label{BFKLBE11}
 &&  \frac{\partial \Phi^{\rm int}\Lb  Y - y, p_{1,T}, p_{2,T}; q_T \Rb}{\partial (Y - y) }\,\,= \,\,\\ 
 && \frac{\bas}{N^2_c - 1}  \int \frac{d^2  Q'_T}{\Lb \Delta\vec{p}_{12,T} - \vec{Q}'_T -\vec{q}_T   \Rb^2} \,\Big\{\Phi\Lb Y - y, p_{2,T}, p_{2,T}; Q'_T\Rb \,-\,\omega_G\Lb\vec{p}_{2,T} \Rb\Phi\Lb Y - y, p_{2,T}, p_{2,T}; q_T\Rb \Big\} \nn\eea  
  The second term in $\{ \dots\}$ stems from the gluon reggeization, 
 in which we neglect $q_T$ in comparison with $p_{i,T}$.      
  Bearing \eq{BFKLBE11} in mind, \eq{EQBFKL3} takes the form:
   
    \bea \label{BFKLBE2}
&&\frac{\partial \Phi\Lb Y - y,p_{1,T}; Y - y,p_{2,T}; q_T\Rb}{\partial \,\Lb Y - y\Rb} \,\,=\,\,\bas \int \frac{d^2 k_T}{2 \pi} \Bigg\{K\Lb p_{1,T}, k_T; q_T\Rb \,\Phi\Lb Y - y,k_{T}; Y - y, p_{2,T}; q_T\Rb
\,+\,\,\Big( 1 \,\leftrightarrow\,2\Big)\Bigg\}\,\nn\\
&&
+\,\frac{\bas}{N^2_c - 1} \Bigg\{
    \int \frac{d^2 Q'_T}{2 \pi} \frac{1}{\Lb \Delta\vec{p}_{12,T} - \vec{Q}'_T -\vec{q}_T   \Rb^2}\, \Phi\Lb Y - y, p_{1,T}; Y - y,  p_{1,T}; Q'_T\Rb\,\,- \,\omega_G\Lb\vec{p}_{2,T} \Rb\Phi\Lb Y - y, p_{2,T}, p_{2,T}; q_T\Rb\nn\\
    &&\hspace{1.5cm} \,\,+\,\,\Big( 1 \,\leftrightarrow\,2\Big) \Bigg\}\,\, \,+\,\,\,\bas^2\,V\Lb \vec{p}_{1,T}, \vec{p}_{2,T}, \vec{q}_T\Rb
\,\,\phi\Lb Y - y', \vec{p}_{1,T} + \vec{p}_{2,T}\Rb
\eea

 We simplify the equation by first neglecting the $q_T$ dependence of
 the BFKL kernel in the first two terms of the R.H.S. of the equation,
 since as has been discussed, $q_T \,\ll\,k_T(p_{i,T})$.
As the second step we go to the impact parameter representation:
\beq \label{BFKLBE3}
\Phi\Lb Y - y,p_{1,T}; Y - y,p_{2,T}; q_T\Rb\,\,=\,\,\int d^2 b\,e^{ i\,\vec{q}_T\,\cdot\,\vec{b}}
\Phi\Lb Y - y,p_{1,T}; Y - y,p_{2,T}; b\Rb
\eeq
 
 \eq{BFKLBE2}  then has the form:
 \bea \label{BFKLBE4}
 &&\frac{\partial \Phi\Lb Y - y,p_{1,T}; Y - y,p_{2,T}; b\Rb}{\partial \,\Lb Y - y\Rb} \,\,=\\
&&\bas \int \frac{d^2 k_T}{2 \pi} \Bigg\{K\Lb p_{1,T}, k_T\Rb \,\Phi\Lb Y - y,k_{T}; Y - y, p_{2,T}; b\Rb
\,+\,K\Lb p_{2,T}, k_T\Rb \,\Phi\Lb Y - y, p_{1,T}; Y - y, k_T; b\Rb\Bigg\}\,\nn\\
&&
+\,\frac{\bas}{N^2_c - 1} \Bigg\{\Bigg(S\Lb b\,p_{1,T}\Rb e^{ i \,\vec{b}\,\cdot\,\Delta\vec{p}_{12,T}}\,-\,S\Lb 0\Rb\Bigg)\Phi\Lb Y - y, p_{1,T}; Y - y,  p_{1,T}; b\Rb\,\,\,\,\,+\,\,\,\,\,\Big( 1 \,\leftrightarrow\,2\Big)\Bigg\}\nn\\
    && \,+\,\,\bas^2\,V\Lb \vec{p}_{1,T}, \vec{p}_{2,T},b\Rb
\,\,\phi\Lb Y - y', \vec{p}_{1,T} + \vec{p}_{2,T}\Rb\nn
\nn
\eea
 
 We need to re-write the kernel $ 1\Big{/}\Lb \Delta\vec{p}_{12,T}
 - \vec{Q}'_T -\vec{q}_T   \Rb^2$ to regularize the infrared
 singularity and to  take into account that $ | \vec{Q}'_T|\, \ll 
\,p_{i,T}$.
 The last constraint was  used in deriving  the equation. We now
 replace the kernel by the expression
  \beq \label{BFKLBE5} 
  \frac{1}{\Lb \Delta\vec{p}_{12,T} - \vec{Q}'_T -\vec{q}_T   \Rb^2  }\,\,\to\,\,\frac{1}{\Lb \Delta\vec{p}_{12,T} - \vec{Q}'_T -\vec{q}_T   \Rb^2 + \mu^2 }\,  \,-\,\,\frac{1}{\Lb \Delta\vec{p}_{12,T} - \vec{Q}'_T -\vec{q}_T   \Rb^2 + p^2_{1,T} }
  \eeq   
  One can see that the term with $\mu^2$ regularizes the infrared
 divergency, and the second term  guarantees that only $ Q'_T$ and
 $q_T$ less than $p_{1,T}$, contribute to the integral.  Calculating
 the Fourier  transform of \eq{BFKLBE5}
  we obtain
  
  \beq \label{BFKLBE6}
  S\Lb b\,p_{1,T}\Rb \,\,=\,\,K_0\Lb \mu\, b\Rb \,-\, K_0\Lb p_{1,T}\,b\Rb  
  \eeq
  
  Since at $b \to 0$ $\Lb b\,p_{1,T}\Rb\,\to \ln\Lb p^2_{1,T} /\mu^2\Rb$,
 the reggeization term with $\omega_G\Lb p_{1,t}\Rb = \ln\Lb p^2_T/\mu^2\Rb$
 cancels the infrared divergency in the  difference $\Bigg(S\Lb b\,p_{1,T}\Rb
 e^{ i \,\vec{b}\,\cdot\,\Delta\vec{p}_{12,T}}\,-\,S\Lb 0\Rb  \Bigg)$.

We first find the solution to the homogeneous equation re-writing it in 
the Mellin transform of \eq{MLN}. It has the form:

\bea 
\label{BFKLBEMLN}
&&\omega\,\Phi\Lb \omega, \gamma_1, \gamma_2; b \Rb\,\,=\\
&&\,\,\bas\,\Lb \chi\Lb\gamma_1\Rb\,+\,\chi\Lb \gamma_2\Rb\Rb\,\Phi\Lb \omega, \gamma_1, \gamma_2; b \Rb\,\,\,\,\,+\,\,\delta \Bigg\{\int^{\epsilon + i \infty}_{\epsilon - i \infty}\frac{d \gamma'_1}{2 \pi i}K\,\Lb \gamma_2,\gamma_1-\gamma'_1; b\Rb  \int^{\epsilon + i \infty}_{\epsilon - i \infty}\frac{d \gamma'}{2 \pi i}
\Phi\Lb \omega, \gamma', \gamma'_1 -   \gamma' ; b \Rb\,\,\nn\\
&&+\,\,\int^{\epsilon + i \infty}_{\epsilon - i \infty}\frac{d \gamma'_1}{2 \pi i}K\,\Lb \gamma_1,\gamma_2-\gamma'_2; b\Rb \int^{\epsilon + i \infty}_{\epsilon - i \infty}\frac{d \gamma'}{2 \pi i}
\Phi\Lb \omega, \gamma', \gamma'_2 -   \gamma'; b  \Rb\Bigg\}\nn
\eea
where $\delta = \bas/(N^2_c -1)$. $K\,\Lb \gamma_2,\gamma_1-\gamma'_1; b\Rb $ is equal to
\bea \label{BFKLBEMLN1}
K\Lb \gamma_2,\gamma_1-\gamma'_1; b\Rb\,\,&=&\,\,\tilde{K}\Lb \gamma_2,\gamma_1-\gamma'_1; b\Rb\,-\,\hat{K}\Lb \gamma_2,\gamma_1-\gamma'_1\Rb;\\
\tilde{K}\Lb \gamma_2,\gamma_1-\gamma'_1; b\Rb \,\,&=&\,\,\int  p_{1,T}\,d p_{1,T} \,  p_{2,T}\,d p_{2,T} \Lb p^2_{1,T}\Rb^{-\gamma_1}
\Lb p^2_{2,T}\Rb^{-\gamma_2} J_0\Lb b\, p_{1,T}\Rb \,J_0\Lb b\, p_{2,T}\Rb\,S\Lb b\, p_{1,T}\Rb\,\Lb  p^2_{1,T}\Rb ^{\gamma'_1 - 1} \nn\\
\hat{K}\Lb \gamma_2,\gamma_1-\gamma'_1\Rb\,\,&=&\,\,\int  p_{1,T}\,d p_{1,T} \,  p_{2,T}\,d p_{2,T} \Lb p^2_{1,T}\Rb^{-\gamma_1}
\Lb p^2_{2,T}\Rb^{-\gamma_2} \,\omega_G\Lb p_{1,T}\Rb\,\Lb  p^2_{1,T}\Rb ^{\gamma'_1 - 1}\nn
\eea

For  $K\Lb \gamma_2,\gamma_1-\gamma'_1; b=0\Rb$ we have 
\beq \label{BFKLBEMLN2}
K\Lb \gamma_2,\gamma_1-\gamma'_1; b=0\Rb\,\,=\,\,-\,\frac{1}{4}\frac{1}{1 - \gamma_2}\,\frac{1}{\Lb\gamma'_1 - \gamma_1\Rb^2}
\eeq
 Integrating over $\gamma'_1$($\gamma'_2$) we get the following equation
\bea 
\label{BFKLBEMLN3}
&&\omega\,\Phi\Lb \omega, \gamma_1, \gamma_2; b \Rb\,\,=\,\,\bas\,\Lb \chi\Lb\gamma_1\Rb\,+\,\chi\Lb \gamma_2\Rb\Rb\,\Phi\Lb \omega, \gamma_1, \gamma_2; b \Rb\,\,\nn\\
&&\,\,\,-\,\,\frac{\delta}{4} \Bigg\{\frac{1}{1 - \gamma_2} \int^{\epsilon + i \infty}_{\epsilon - i \infty}\frac{d \gamma'}{2 \pi i}
\Phi'_{\gamma_1}\Lb \omega, \gamma', \gamma_1 -   \gamma' ; b \Rb\,\,+\,\,\frac{1}{1 - \gamma_1}\int^{\epsilon + i \infty}_{\epsilon - i \infty}\frac{d \gamma'}{2 \pi i}
\Phi'_{\gamma_2}\Lb \omega, \gamma', \gamma_2 -   \gamma'; b  \Rb\Bigg\}\nn
\eea

We solve this equation using the iteration procedure with respect to the
small parameter $\delta$, assuming that the solution without the 
interference
 term is equal to
\beq \label{BESOL1}
\Phi^{(0)}\Lb \omega, \gamma_1, \gamma_2\Rb\,\,=\,\,\frac{1}{ \omega\,-\,\bas\Lb \chi\Lb \gamma_1\Rb + \chi\Lb \gamma_2\Rb\Rb}
\eeq

Plugging this solution in \eq{BFKLBEMLN},  we obtain the 
following equation
 for the spectrum of the homogeneous equation:
\beq \label{BESOL2}
1\,\,=\,\,-\,\frac{\delta}{4}\,\Bigg\{\frac{ 1}{1 - \gamma_2}\int^{\epsilon + i \infty}_{\epsilon - i \infty}\frac{d \gamma'}{2 \pi i}
\Phi^{(0)'}_{\gamma_1}\Lb \omega, \gamma', \gamma_1 -   \gamma' \Rb\,\,+\,\,\frac{1}{1 - \gamma_1}  \int^{\epsilon + i \infty}_{\epsilon - i \infty}\frac{d \gamma'}{2 \pi i}
\Phi^{(0)'}_{\gamma_2}\Lb \omega, \gamma', \gamma_2 -   \gamma' \Rb\Bigg\}
\eeq
In  general for  the BFKL kernel (see \eq{KER})  we cannot integrate  
the
 integral analytically. Instead, we use the diffusion approximation to
 obtain the analytical result:
\bea \label{BESOL3}
 \int^{\epsilon + i \infty}_{\epsilon - i \infty}\frac{d \gamma'}{2 \pi i}\Phi^{(0)}\Lb \omega, \gamma', \gamma -   \gamma' \Rb\,\,&=&\,\, \int^{\epsilon + i \infty}_{\epsilon - i \infty}\frac{d \gamma'}{2 \pi i}\frac{1}{\omega - \bas \underbrace{ \Lb 2 \omega_0 + D\Lb  \Lb \gamma' -\h\Rb^2 + 
 \Lb \gamma - \gamma' - \h\Rb^2\Rb\Rb}_{\mbox{diffusion approximation}}}\\
 & &\nn\\
  &=& \h \frac{1}{\sqrt{2 \bas  D \Lb\omega \,-\,2 \bas \Lb\omega_0 + D\Lb \frac{\gamma}{2} - \h\Rb^2\Rb\Rb}}\,=\, \h \frac{1}{\sqrt{2 \bas D \Lb\omega \,-\,2 \bas \chi_{\rm diff. ~app.}\Lb \h \gamma\Rb\Rb}} \nn\eea
where $ \chi_{\rm diff. ~app.}$ is the BFKL kernel in the diffusion
 approximation of \eq{CHI}.

Plugging \eq{BESOL3} into \eq{BESOL2},  we reduce this equation to the
 form for $\gamma_1=\gamma_2$
\beq \label{BESOL4}
1\,\,=\,\,\,\frac{\delta}{4\sqrt{2}}\sqrt{\bas D}\frac{1}{ \Lb\omega \,-\,2 \bas \Lb\omega_0 + D\Lb \frac{\gamma}{2} - \h\Rb^2\Rb\Rb^{3/2}}
\eeq

Searching for the solution with the new intercept $\omega - 2
 \bas \omega_0 = \omega^{(1)}\Lb \gamma_1,\gamma_2\Rb$, we obtain
 the solution for $\gamma_1=\gamma_2 \to 1$.
\beq \label{BESOL5}
\omega^{(1)}\Lb \gamma_1,\gamma_2\Rb\,\,=\,\Lb \frac{\delta}{4\sqrt{2}}\sqrt{\bas D}\Rb^{2/3}\,\,=\,\,\frac{\bas}{\Lb N^2_c -1\Rb^{2/3}}\,\Lb \frac{D}{32}\Rb^{1/3}\,\,\approx\,\, 0.8\,\frac{\bas}{\Lb N^2-1\Rb^{2/3}}\,=\,0.2 \,\bas ~~\mbox{for} N_c=3
\eeq

Therefore, we see that the value  of the  intercept for the double parton
  density ($Y  - y$) dependence,  is larger than  that  for the product
 of single parton densities (see \eq{I2}), which is equal to

\beq \label{BESOL5}
\Phi\Lb Y - y,p_{1,T}; Y - y,p_{2,T}; q_T\Rb\,\propto\,e^{ \Delta_2 \Lb Y - y\Rb}; ~~~~~\mbox{with}
~~~\Delta_2 \,=\,2\bas \omega_0 \,\,+\,\, \frac{\bas}{\Lb N^2_c -1\Rb^{2/3}}\,\Lb \frac{D}{32}\Rb^{1/3}\eeq

Therefore, the difference $\Delta_2 - 2 \bas \omega_0$ turns out to be
 proportional to $1/\Lb N^2_c - 1\Rb^{2/3}$,  which is a small number at
 large $N_c$. However, for $N_c=3$ (see \eq{BESOL5})  $\Delta_2 - 2 \bas
 \omega_0 \approx 0.2 \,\bas \approx 0.07 \bas \omega_0$. This value of
 the correction leads to an effect of the order of 1 at $Y \approx 20$.

 We can calculate the corrections of the order of $\delta^2$ to the 
intercept,
 using for the iteration 
 $\Phi^{(1)}\Lb \omega, \gamma_1, \gamma_2\Rb $  the form
 \beq \label{BESOL7}
\Phi^{(1)}\Lb \omega, \gamma_1, \gamma_2\Rb\,\,=\,\,\frac{1}{ \omega\,-\,2 \omega_0 \,-\,D\Lb \Lb \gamma_1 - \h\Rb^2 + \Lb \gamma_2 - \h\Rb^2\Rb  \,\,-\,\,\omega^{(1)}\Lb \gamma_1, \gamma_2\Rb}
\eeq
 
 However, we expect  negligible values for this correction, and  we 
proceed
 to find the solution of 
 \eq{BFKLBE2} using \eq{BESOL7}, as the solution for the homogeneous 
equation.

  \section{High energy behaviour of the double gluon densities}
    
   Using the result of \eq{BESOL7} we can write  \eq{EQMLN} 
in the following form:
    \beq 
\label{BEEQMLN}
\omega\,\Phi\Lb \omega, \gamma_1, \gamma_2\Rb\,\,=\,\,\bas\,\Lb \chi\Lb\gamma_1\Rb\,+\,\chi\Lb \gamma_2\Rb\,\,+\,\,\omega^{(1)}\Lb \gamma_1, \gamma_2\Rb\Rb\,\Phi\Lb \omega, \gamma_1, \gamma_2\Rb\,\,+\,\,H\Lb \gamma_1, \gamma_2\Rb \phi\Lb \gamma_1 + \gamma_2\Rb
\eeq    
 \eq{BEEQMLN} generates the following particular solution, which is a
 direct generalization of \eq{SOL2}:
 \bea \label{BESOL8}
&&\Phi_{\rm part}\Lb Y - y,p_{1,T}; Y - y,p_{2,T}; q_T\Rb\,\,=\,\,\int^{\epsilon + i \infty}_{\epsilon - i\infty}\frac{d \gamma_1}{2 \pi i}\int^{\epsilon + i \infty}_{\epsilon - i\infty}\frac{d \gamma_2}{2 \pi i}\\&&\,e^{ \gamma_1 \xi_1 \,+\,\gamma_2 \xi_2} \frac{\bas\,H\Lb \gamma_1,\gamma_2\Rb \,\phi_{\rm in}\Lb \gamma_1 + \gamma_2\Rb}{\omega\Lb \gamma_1\Rb\,+\,\omega\Lb \gamma_2\Rb\,+\,\omega^{(1)}\Lb \gamma_1, \gamma_2\Rb\,-\,\omega\Lb \gamma_1 + \gamma_2\Rb}\Bigg\{ e^{\Lb\omega\Lb \gamma_1\Rb\,+\,\omega\Lb \gamma_2\Rb\,\,+\,\,\omega^{(1)}\Lb \gamma_1, \gamma_2\Rb\Rb\,\Lb Y - y\Rb}
\,\,-\,\,e^{ \bas  \chi\Lb \gamma_1 + \gamma_2\Rb\,\Lb Y - y\Rb}\Bigg\}\nn
\eea 
   The general solution can be written in the same form as \eq{SOL3}
 leading to the following expression
  \bea \label{BESOL9}
&&\Phi\Lb Y - y,p_{1,T}; Y - y,p_{2,T}; q_T\Rb\,\,=\,\,\,
\int^{\epsilon + i \infty}_{\epsilon - i\infty}\frac{d \gamma_1}{2 \pi i}\int^{\epsilon + i \infty}_{\epsilon - i\infty}\frac{d \gamma_2}{2 \pi i}\,e^{ \gamma_1 \xi_1 \,+\,\gamma_2 \xi_2} \\
&&\Bigg\{\Bigg[\frac{\bas^2 H\Lb \gamma_1,\gamma_2\Rb \,\phi_{\rm in}\Lb \gamma_1 + \gamma_2\Rb}{\Lb \omega\Lb \gamma_1\Rb\,+\,\omega\Lb \gamma_2\Rb\,+\omega^{(1)}\Lb \gamma_1, \gamma_2\Rb\,-\,\omega\Lb \gamma_1 + \gamma_2\Rb\Rb}\,+\,
\Phi_{\rm in}\Lb \gamma_1, \gamma_2\Rb\Bigg] e^{\Lb \omega\Lb \gamma_1\Rb\,+\,\omega\Lb \gamma_2\Rb\,+\,\omega^{(1)}\Lb\gamma_1, \gamma_2\Rb\Rb\,\Lb Y - y\Rb}\nn\\
&&
\,\,-\,\,\frac{\bas^2 H\Lb \gamma_1,\gamma_2\Rb \,\phi_{\rm in}\Lb \gamma_1 + \gamma_2\Rb}{\Lb \omega\Lb \gamma_1\Rb\,+\,\omega\Lb \gamma_2\Rb\,+\omega^{(1)}\Lb \gamma_1, \gamma_2\Rb\,-\,\omega\Lb \gamma_1 + \gamma_2\Rb \Rb}e^{ \bas  \chi\Lb \gamma_1 + \gamma_2\Rb\,\Lb Y - y\Rb}\Bigg\}\nn\eea 
   
   As have been discussed previously, we can use the method of steepest 
descend to  evaluate
the integrals over $\gamma_1$ and $\gamma_2$ in the first term of
 \eq{BESOL9}. To do this we need to know the dependence of
 $\omega^{(1)}\Lb \gamma_1,\gamma_2\Rb$ in $\gamma_1$ and $\gamma_2$.
 Solving \eq{BESOL4} we see that 
\beq \label{BESOL90}
\omega^{(1)}\Lb \gamma,\gamma\Rb\,\,=\,\,\omega^{(1)}_0 \,-\,\,\frac{3}{2}\,\bas\,D\, \gamma\,\Lb  \gamma- \frac{2}{3}\Rb
\eeq
where $\omega^{(1)}_0 $ is given by \eq{BESOL5}.

 The values of the saddle point turn out to be close to $\h$  for
 both $\gamma$'s. In the vicinity   $\gamma_1 \to \h$ and $\gamma_2 \to \h$ we
 have the following expansion
\beq \label{BESOL10}
 \omega\Lb \gamma_1\Rb\,+\,\omega\Lb \gamma_2\Rb\,+\,\omega^{(1)}\Lb\gamma_1, \gamma_2\Rb\,\,=\,\, 2 \bas \omega_0 \,+\,\omega^{(1)}_0 \,+\,  \frac{\bas\,D}{4} \,  \Lb \Lb1 -  \gamma_1 \Rb^2 + \Lb 1 - \gamma_2\Rb^2 \Rb 
   \eeq
From \eq{BESOL10} we obtain that  the values of  $\gamma$'s at the
 saddle point are the following:
\beq \label{SPGA}
\gamma^{\rm SP}_1\,\,=\,\,1 - \frac{2\,\xi_1}{\bas\, D\, \Lb Y - y\Rb};~~~~~~~\gamma^{\rm SP}_2\,\,=\,\,1 - \frac{2\,\xi_2}{ \bas D \Lb Y - y\Rb};
\eeq

After integration over $\gamma_1$ and $\gamma_2$ in vicinities of these 
saddle points we obtain the contribution:
\bea \label{SOL4}
&&\Phi\Lb Y - y,p_{1,T}; Y - y,p_{2,T}; q_T\Rb\,\,=\\
&&\,\,\frac{1}{4 \pi}\frac{2}{ \bas D \Lb Y - y\Rb}\Bigg[ \dots\Bigg]_{\gamma_1 = \gamma^{\rm SP}_1; \gamma_2 = \gamma^{\rm SP}_2}  \!\!\!\!\!\!\!\!\exp\Lb 2 \bas  \, \Delta_2 \Lb Y - y\Rb\,  \,-\,2\,\frac{\xi^2_1 + \xi_2^2}{ \bas D \Lb Y - y\Rb}
\Rb\nn
\eea
One can see that this contribution is not  proportional
 to $\phi\Lb Y - y, \xi_1\Rb\,\phi\Lb Y - y, \xi_2\Rb$, 
 and contradicts \eq{I2}. It should be stressed that \eq{I2}
 violates both the $Y - y$ dependence due to the intercept
 $\Delta_2\, > \,2 \bas \omega_0$, and the $\xi_1$($\xi_2$)
 dependence (compare this equation with \eq{SOL4}). Note that
 the change in the $\xi$ shape of the distribution, has no
 suppression of $1/\Lb N^2_c-1\Rb$.

    \vspace{2cm}

  \section{Conclusions}
 In this paper we found that in the BFKL evolution, the
 Bose-Einstein enhancement leads to a faster increase of the
 double parton densities than the product    of two single parton
 distributions.  This effect has been discussed by us for the DGLAP 
evolution in the region of low $x$\cite{GOLELAST}. On the qualitative
 level, the DGLAP and BFKL evolution lead to large correlations at high 
energies,  due to the correlations of the identical gluons. 
It should be noted that {\it all }  $1/(N^2_c-1)$ corrections in the
 double gluon densities stem from the Bose-Einstein enhancement.

   The BFKL evolution generates the power dependence on $x$ (
 $\Phi \propto \Lb 1/x\Rb^{\Delta_2}$ with $\Delta_2 - 2 \omega_0 > 0$,
 where $\omega_0$ is the intercept of the BFKL Pomeron, this difference
 turns out to be  numerically small, since it is proportional to
 $\bas/\Lb N^2_c - 1\Rb^{2/3}$.

  In particular, these correlations clarify the physical meaning
 of the increase of the anomalous dimension of the twist four operator
 that has been discussed in Refs. \cite{BART1,BART2,LET2,LLS,LALE,BARY}.
 It should be stressed that we obtain the 
  intercept for the double gluon density is proportion; to $1/\Lb N^2_c -
 1\Rb^{2/3}$ which is quite different from the twist four intercept, which
 is proportional to $1/\Lb N^2_c - 1\Rb^2$.
   However,  in the case of the anomalous
 dimensions,  corrections other than the Bose-Einstein enhancement, of
 the order of $1/(N^2_c-1)$, contribute so making the calculation of the
 energy behavior of the twist four operator to be a  different problem.

  We view this paper as the next step in understanding  the role of
 the identical parton correlations in the parton evolution. The next
project 
that we plan to consider, is  to include the identical parton correlations 
in the DGLAP evolution at finite $x$. We believe that the most
 interesting question in his part of the program is to include
 the Pauli blocking for quarks and antiquarks (see Ref.\cite{ALAM}).

  \section{Acknowledgements}
   We thank our colleagues at Tel Aviv university and UTFSM for
 encouraging discussions. Our special thanks go to  
 Alex Kovner and Misha Lublinsky for elucidating discussions on the
 role of the Bose - Einstein correlation in the CGC  effective theory.
 
  This research was supported by the BSF grant   2012124, by 
   Proyecto Basal FB 0821(Chile),  Fondecyt (Chile) grant  
 1180118 and by   CONICYT grant PIA ACT1406.  
 
 ~

\end{document}